\PassOptionsToPackage{hypertexnames=false}{hyperref}

\documentclass[pdflatex,sn-basic]{sn-jnl}

\usepackage{graphicx}
\usepackage{xurl}
\usepackage{float}
\usepackage{amsmath,amssymb,amsfonts}
\usepackage{mathtools}
\usepackage{nicefrac}
\usepackage[group-minimum-digits=5]{siunitx}
\usepackage{booktabs}
\usepackage{tabularx}
\usepackage{listings}
\usepackage{textcomp}
\usepackage{pifont}
\usepackage[table]{xcolor}
\usepackage{tikz}
\let\orcidlogo\relax
\usepackage{orcidlink}
\usetikzlibrary{positioning,patterns}

\usepackage{pgfplots}
\pgfplotsset{compat=1.18}

\definecolor{histAllatori}{HTML}{F4C27C}
\definecolor{histProGuard}{HTML}{AECFEF}
\definecolor{histYGuard}{HTML}{D1CAF9}
\definecolor{histHatch}{HTML}{969696}
\definecolor{histGrid}{HTML}{DDDDDD}
\definecolor{histFrame}{HTML}{000000}
\definecolor{histTick}{HTML}{777777}
\definecolor{aigenerated}{HTML}{895129}
\usepackage[most]{tcolorbox}
\usepackage{ifthen}
\usepackage{xspace}

\definecolor{simoncomment}{HTML}{6A1B9A}
\definecolor{valeriocomment}{HTML}{C62828}
\definecolor{danielecomment}{HTML}{2E7D32}
\definecolor{giuseppecomment}{HTML}{1565C0}
\definecolor{leonardocomment}{HTML}{EF6C00}
\definecolor{pascalcomment}{HTML}{00838F}

\newboolean{showcomments}
\setboolean{showcomments}{false}
\ifthenelse{\boolean{showcomments}}{
  \newcommand{\mynote}[3]{
    \fbox{\bfseries\sffamily\footnotesize#1}
    {\small$\blacktriangleright$
     \textsf{\emph{\color{#3}{#2}}}$\blacktriangleleft$
     }
  }
}{
  \newcommand{\mynote}[3]{}
}

\newcommand{\ie}{\textit{i.e.}\xspace}
\newcommand{\eg}{\textit{e.g.}\xspace}
\newcommand{\etc}{\textit{etc.}\xspace}
\newcommand{\quotes}[1]{``#1''}

\definecolor{blueboxbackground}{HTML}{EAF3FF}
\definecolor{blueboxborder}{HTML}{2F6FDB}
\definecolor{obfuscatedlexeme}{HTML}{FFA3A3}
\definecolor{recoveredlexeme}{HTML}{9CFFA9}
\definecolor{groundtruthlexeme}{HTML}{9CFFFD}
\newcommand{\obfhl}[1]{{\setlength{\fboxsep}{0.4pt}\smash{\colorbox{obfuscatedlexeme}{\rule[-0.88pt]{0pt}{4.96pt}\textcolor{black}{\ttfamily #1}}}}}
\newcommand{\recoveredhl}[1]{{\setlength{\fboxsep}{0.4pt}\smash{\colorbox{recoveredlexeme}{\rule[-0.88pt]{0pt}{4.96pt}\textcolor{black}{\ttfamily #1}}}}}
\newcommand{\groundtruthhl}[1]{{\setlength{\fboxsep}{0.4pt}\smash{\colorbox{groundtruthlexeme}{\rule[-0.88pt]{0pt}{4.96pt}\textcolor{black}{\ttfamily #1}}}}}
\newcommand{\bluebox}[1]{%
  \par\vspace{2pt}%
  \noindent\begingroup
  \setlength{\fboxsep}{0.5em}%
  \fcolorbox{blueboxborder}{blueboxbackground}{%
    \parbox{0.94\linewidth}{#1}%
  }%
  \endgroup
  \par\vspace{2pt}%
}
\makeatletter
\long\def\@figurecaption#1#2{%
  {\figurecaptionfont\centering{\bfseries#1}\hskip.7em#2\par}%
}
\makeatother

\usepackage{etoolbox}
\hypersetup{colorlinks=true, citecolor=blue,
  pdftitle={More Context, Mixed Results: Evaluating LLMs for Identifier Recovery and Code Deobfuscation},
  pdfauthor={Simon Queyrut, Valerio Schiavoni, Giuseppe Antonio Di Luna, Daniele Cono D'Elia, Leonardo Querzoni, Pascal Felber}}

\makeatletter

\patchcmd{\NAT@citex}
  {\@citea\NAT@hyper@{%
     \NAT@nmfmt{\NAT@nm}%
     \hyper@natlinkbreak{%
       \NAT@aysep\NAT@spacechar}{\@citeb\@extra@b@citeb}%
     \NAT@date}}
  {\@citea\NAT@nmfmt{\NAT@nm}%
   \NAT@aysep\NAT@spacechar\NAT@hyper@{\NAT@date}}
  {}
  {\PackageWarning{yearlinks}{Parenthetical citation patch failed}}

\patchcmd{\NAT@citex}
  {\@citea\NAT@hyper@{%
     \NAT@nmfmt{\NAT@nm}%
     \hyper@natlinkbreak{%
       \NAT@spacechar\NAT@@open\if*#1*\else#1\NAT@spacechar\fi}%
       {\@citeb\@extra@b@citeb}%
     \NAT@date}}
  {\@citea\NAT@nmfmt{\NAT@nm}%
   \NAT@spacechar\NAT@@open\if*#1*\else#1\NAT@spacechar\fi
   \NAT@hyper@{\NAT@date}}
  {}
  {\PackageWarning{yearlinks}{Textual citation patch failed}}

\patchcmd{\HyRef@autosetref}
  {#3{%
      \expandafter\@fifthoffive#1\@empty\@empty\@empty
    }{%
      \expandafter\@fourthoffive#1\@empty\@empty\@empty
    }{%
      \HyRef@currentHtag
      \expandafter\@firstoffive#1\@empty\@empty\@empty
      \null
    }}
  {\HyRef@currentHtag
   #3{%
      \expandafter\@fifthoffive#1\@empty\@empty\@empty
    }{%
      \expandafter\@fourthoffive#1\@empty\@empty\@empty
    }{%
      \expandafter\@firstoffive#1\@empty\@empty\@empty
      \null
    }}
  {}
  {\PackageWarning{numberlinks}{Autoref number-only link patch failed}}

\makeatother

\let\cite\citep

\makeatletter
\patchcmd{\@maketitle}{\else\vskip21pt\fi}{\else\vskip0pt\fi}
  {}{\PackageWarning{firstpage}{Title top-spacing patch failed}}
\patchcmd{\@maketitle}{\vskip20pt\nointerlineskip}{\vskip0pt\nointerlineskip}
  {}{\PackageWarning{firstpage}{Title heading-spacing patch failed}}
\patchcmd{\@maketitle}{\vskip24pt}{\vskip10pt}
  {}{\PackageWarning{firstpage}{Contact top-spacing patch failed}}
\patchcmd{\@maketitle}{\vskip24pt}{\vskip10pt}
  {}{\PackageWarning{firstpage}{Contact bottom-spacing patch failed}}
\patchcmd{\@maketitle}{\vskip36pt\vskip0pt}{\vskip18pt\vskip0pt}
  {}{\PackageWarning{firstpage}{Title bottom-spacing patch failed}}
\makeatother

\begin{document}

\title[LLMs for Identifier Recovery and Code Deobfuscation]{More Context, Mixed Results: Evaluating LLMs for Identifier Recovery and Code Deobfuscation}

\author*[1]{\fnm{Simon} \sur{Queyrut}\,\orcidlink{0000-0002-1354-9604}}
\email{\href{mailto:simon.queyrut@unine.ch}{simon.queyrut@unine.ch}}
\author[1]{\fnm{Valerio} \sur{Schiavoni}\,\orcidlink{0000-0003-1493-6603}}
\email{\href{mailto:valerio.schiavoni@unine.ch}{valerio.schiavoni@unine.ch}}
\author[2]{\fnm{Giuseppe Antonio} \sur{Di Luna}\,\orcidlink{0000-0002-7150-0972}}
\email{\href{mailto:diluna@diag.uniroma1.it}{diluna@diag.uniroma1.it}}
\author[2]{\fnm{Daniele Cono} \sur{D'Elia}\,\orcidlink{0000-0003-4358-976X}}
\email{\href{mailto:delia@diag.uniroma1.it}{delia@diag.uniroma1.it}}
\author[2]{\fnm{Leonardo} \sur{Querzoni}\,\orcidlink{0000-0002-8711-4216}}
\email{\href{mailto:querzoni@diag.uniroma1.it}{querzoni@diag.uniroma1.it}}
\author[1]{\fnm{Pascal} \sur{Felber}\,\orcidlink{0000-0003-1574-6721}}
\email{\href{mailto:pascal.felber@unine.ch}{pascal.felber@unine.ch}}

\affil[1]{\orgdiv{Computer Science Department}, \orgname{University of Neuch\^atel}, \orgaddress{\city{Neuch\^atel}, \country{Switzerland}}}
\affil[2]{\orgdiv{Department of Computer, Control and Management Engineering}, \orgname{Sapienza University of Rome}, \orgaddress{\city{Rome}, \country{Italy}}}

\abstract{\unboldmath
Source code is often obfuscated through identifier renaming, incentivizing reverse engineers to develop techniques for recovering meaningful names.
To assess how LLMs compare with these techniques, we evaluate 4 traditional baselines and 3 LLMs (Qwen, OLMo and DeepSeek) on identifier detection and semantic recovery across 51 recent Java projects, 3 obfuscators, and 6 context/retrieval settings.
LLMs generally outperform traditional techniques on both tasks, but benefit from different forms of context and settings:
agentic retrieval yields the strongest detection results, reaching 91.56\% $F_1$, while enclosing-class context generally improves semantic recovery at a higher token cost.
Despite these gains, exact recovery peaks at 16.42\% overall (under one-hop context) and 32.93\% for field names (under repository retrieval), which are the easiest to recover.
Class identifiers are the hardest to detect, with a mean detection $F_1$ of 48.48\% across models, settings and obfuscators.
More context is not always beneficial: larger methods hinder semantic recovery.
Beyond identifier renaming, our experiments show that control-flow obfuscation degrades both tasks more than string encryption.
Our findings provide useful ad hoc insights for reverse engineers choosing models and context strategies to deobfuscate code snippets.
}

\maketitle

\begin{figure}[!t]
\centering
\lstset{
  basicstyle=\ttfamily\fontsize{4.8}{6}\selectfont,
  columns=fullflexible,
  keepspaces=true,
  showstringspaces=false,
  escapeinside={(*@}{@*)},
  numbers=left,
  numberstyle=\sffamily\fontsize{4}{6}\selectfont\color{black!35},
  numbersep=3pt,
  stepnumber=1,
  xleftmargin=1cm,
  aboveskip=6pt,
  belowskip=0pt
}
\begin{minipage}[t]{0.322\linewidth}
\centering\makebox[\linewidth][c]{\hspace*{-0.25cm}\sffamily\bfseries\fontsize{7}{8}\selectfont Obfuscated}\par
\begin{lstlisting}
public boolean (*@\obfhl{g}@*)() {
 try {
  URL uRL = this.(*@\obfhl{j}@*)();
  if (ResourceUtils
      .isFileURL((URL) uRL)) {
   File file = this.(*@\obfhl{i}@*)();
   return file.canRead()
       && !file.isDirectory();
  }
  return true;
 } catch (IOException
          iOException) {
  (*@\obfhl{a}@*).debug("STR", (Object) this.(*@\obfhl{e}@*),
      (Object) iOException);
  return false;
 }
}
\end{lstlisting}
\end{minipage}\hfill
\begin{minipage}[t]{0.322\linewidth}
\centering\makebox[\linewidth][c]{\hspace*{0.15cm}\sffamily\bfseries\fontsize{7}{8}\selectfont Reverse-engineered}\par
\begin{lstlisting}
public boolean (*@\recoveredhl{isReadable}@*)() {
 try {
  URL uRL = this.(*@\recoveredhl{getResource}@*)();
  if (ResourceUtils
      .isFileURL((URL) uRL)) {
   File file = this.(*@\recoveredhl{getFile}@*)();
   return file.canRead()
       && !file.isDirectory();
  }
  return true;
 } catch (IOException
          iOException) {
  (*@\recoveredhl{logger}@*).debug("STR",
      (Object) this.(*@\recoveredhl{path}@*),
      (Object) iOException);
  return false;
 }
}
\end{lstlisting}
\end{minipage}\hfill
\begin{minipage}[t]{0.322\linewidth}
\centering\makebox[\linewidth][c]{\hspace*{-0.25cm}\sffamily\bfseries\fontsize{7}{8}\selectfont Ground truth}\par
\begin{lstlisting}
public boolean (*@\groundtruthhl{isReadable}@*)() {
 try {
  URL uRL = this.(*@\groundtruthhl{getURL}@*)();
  if (ResourceUtils
      .isFileURL((URL) uRL)) {
   File file = this.(*@\groundtruthhl{getFile}@*)();
   return file.canRead()
       && !file.isDirectory();
  }
  return true;
 } catch (IOException
          iOException) {
  (*@\groundtruthhl{log}@*).debug("STR",
      (Object) this.(*@\groundtruthhl{path}@*),
      (Object) iOException);
  return false;
 }
}
\end{lstlisting}
\end{minipage}
\caption{A code snippet obfuscated by Allatori~\cite{allatoriDocumentation} and recovered by one of our evaluated models (Qwen).
Identifiers are exact matches or semantically plausible alternatives.
}
\label{fig:semantic-recovery-example}
\end{figure}

\section{Introduction}
\label{sec:introduction}

Developers and vendors distribute applications that run on devices beyond their control. 
This can expose business logic and proprietary algorithms, facilitate plagiarism and intellectual-property theft, and help bypass licensing mechanisms.
Such threats occur at a large scale: repackaged applications (\ie, existing apps that third parties modify and redistribute) account for 13\% of some third-party markets and that an important fraction of malware consists of modified versions of otherwise benign applications~\cite{DBLP:conf/acsac/WermkeHARTF18}. 
The problem is especially prevalent in Java and Android applications (APKs), which are distributed as structured bytecode and can often be decompiled into readable Java-like code. 
To mitigate these outcomes, developers endeavor to protect their product through schemes that can be broadly grouped into three categories: \textit{(a)} software-based obfuscation, \textit{(b)} encryption, and \textit{(c)} tamper-resistant hardware protection such as Trusted Execution Environments~\cite{banescu2018tutorial}. 
The latter remains comparatively less widely deployed and requires significant engineering effort.
Encryption makes the source code unreadable, but the instructions must eventually be made available in plain form and the decrypted runtime memory image can be exploited.
Fully homomorphic encryption is a particular case where the program stays encrypted during execution, but evaluating encrypted instructions and memory accesses incurs important computational overhead, and the outputs still require decryption, which makes it impractical for general-purpose software protection~\cite{MatsuokaEtAl2021VSP}.
Clearly, \textit{(a)} and \textit{(b)} can be compounded and work complementarily, but only \textit{(a)} reasonably addresses the \quotes{last-mile} problem of runtime transparency. 
Even if the program is decrypted, obfuscation provides a protection that hinders the reverse-engineering process, for instance by transforming the control or data flow, encrypting strings or renaming identifiers \etc 
This work focuses on identifier renaming~\cite{CollbergNagra2009SurreptitiousSoftware,CeccatoEtAl2009Obfuscation,banescu2018tutorial}, a technique that preserves the program's functionality while replacing informative identifiers with meaningless names.
Large-scale APK studies show identifier renaming is a popular protection technique~\cite{DongEtAl18}.

For example, consider in \autoref{fig:semantic-recovery-example} an obfuscated Java method \texttt{g()} that obtains a \texttt{URL}, checks whether the corresponding file is readable, and logs an \texttt{IOException} otherwise. 
When provided with this method together with its enclosing class and direct caller and callee methods, Qwen~\cite{qwen3}, a large language model (LLM) with reasoning disabled, produced the middle snippet in \autoref{fig:semantic-recovery-example}.
This illustrates the two tasks applied in sequence: the model first detected the five obfuscated identifiers and then proposed replacement names, recovering three of the original identifiers exactly.
The exact recoveries make the method easier to understand, while the two plausible near-misses highlight a limitation of evaluating exact matches only.
In the recovered snippet, \texttt{getResource} (line~3) replaces \texttt{getURL}, and \texttt{logger} (line~13) replaces \texttt{log}; both alternatives reasonably convey the identifiers' roles despite not being strictly identical to the original names.   
This example motivates our study of how progressively richer program context affects both exact identifier recovery and semantic similarity.

Before the advent of LLMs, identifier recovery was typically addressed with static analysis and information retrieval techniques.
We will show during our experiments
that they are remarkably ineffective at this task and that LLMs largely outperform them in most of the considered settings.
Research on LLM-based deobfuscation is nonetheless nascent.
The closest previous attempts at this task present important caveats: VarBERT~\cite{DBLP:conf/sp/PalBBDNBGSASDBW24} operates only at the method level and does not allow for additional context.
Moreover, candidate variables are identified and masked by preprocessing before being passed to the model: this means that VarBERT does not need to determine which identifiers require recovery.
This assumption appears natural for variables reconstructed from stripped binaries, but it does not directly apply to deliberate source-code obfuscation, where meaningful and obfuscated identifiers may coexist.
A similar pitfall is found in GENNM~\cite{DBLP:conf/ndss/Xu0SH0Y0X00025}, which operates exclusively on stripped binaries: it will perform recovery only over a known set of artifacts of the stripping process.
We mention very recent work in ObscuraCoder~\cite{DBLP:conf/iclr/PaulYGKG25}, which uses obfuscation and deobfuscation during pretraining to improve downstream coding capabilities: they do not evaluate identifier recovery performance or discuss the influence of varying context/retrieval techniques.

Our work fills this gap by evaluating LLMs on identifier recovery in a setting that is closer to real-world reverse-engineering scenarios. 
Specifically, we tackle the following research questions:

\bluebox{\textbf{\hypertarget{rq:1}{\hyperlink{aq:1}{RQ1}}:} How do LLMs perform at identifier recovery against traditional baseline approaches?}
\bluebox{\textbf{\hypertarget{rq:2}{\hyperlink{aq:2}{RQ2}}:} Are some identifier types harder to recover than others?}
\bluebox{\textbf{\hypertarget{rq:3}{\hyperlink{aq:3}{RQ3}}:} How does identifier-recovery accuracy change with increasing amounts of obfuscated context and varying retrieval strategies?} 
\bluebox{\textbf{\hypertarget{rq:4}{\hyperlink{aq:4}{RQ4}}:} How robust is identifier recovery to changes in the inference setup and compounding obfuscations?}

Our study makes three main contributions. 
First, it provides an extensive side-by-side comparison of traditional identifier recovery techniques and LLM-based approaches to establish a new state of the art for this task. %
Second, we uncover useful insights about how different types of context and retrieval strategies affect the LLMs' output, including trade-off analysis between raw performance and token cost.
Lastly, we present an ablation study against our design choices and shed light on the robustness of LLM-based identifier deobfuscation.

This paper is organized as follows.
We provide background information on code obfuscation and LLMs in \autoref{sec:background}.
\autoref{sec:design} introduces the broad design choices for this deobfuscation study and \autoref{sec:experimental-setup} exposes their practical implementation details.
We share our results in \autoref{sec:results} and in \autoref{sec:ablation}, an ablation study; we reflect on these findings in \autoref{sec:discussion}.
We show in \autoref{sec:related-work} how some previous works relate to our paper, before concluding with \autoref{sec:conclusion}.

\section{Background}
\label{sec:background}

We provide here high-level information on Java obfuscation and deobfuscation (\autoref{ssec:javaobfdeobf}), and on large language models (LLMs) (\autoref{ssec:llm}), specifically for their relevance to the chosen deobfuscation task, \ie identifier recovery.

\subsection{Java Obfuscation and Deobfuscation}\label{ssec:javaobfdeobf}

\textbf{Java obfuscation.}
Code obfuscation transforms a program into a functionally equivalent but harder-to-understand representation to increase the effort required for reverse engineering~\cite{DBLP:journals/tse/CollbergT02}.
ProGuard and yGuard are the two best-known established open-source bytecode obfuscators for general-purpose Java applications.
ProGuard shrinks and optimizes code,  replacing class, method, and field names with meaningless identifiers~\cite{proguard}.
yGuard offers configurable renaming for packages, classes, methods, and fields, alongside code shrinking~\cite{yguard}. 
It is worth noting that while outside the scope of this work, Android developers use R8 to shrink APK code and minify identifiers during the build~\cite{r8}. 

Commercial Java obfuscators, such as Allatori, DashO and Zelix KlassMaster~\cite{dasho,zelix}, provide additional transformations, including string encryption and control-flow obfuscation: we explore compounding such obfuscations later in our study.

Recently, ProGuard and Allatori were identified as being the most frequent obfuscation tools among a large sample of Android apps (Java-based), while identifier renaming was independently identified as the most prevalent obfuscation technique~\cite{niroshan2025empirical}.
While we are aware of this work, it did not undergo peer-review; hence, we conducted a cross-sectional partial replication, reported in Appendix~\ref{app:obfuscator-replication}, which qualitatively supported our focus on ProGuard and Allatori.

\textbf{Identifier obfuscation in Java.}
In Java, identifier obfuscation is a transformation that can target packages, classes, fields, and methods, as well as parameter and local-variable names when these are retained in class-file metadata.
More aggressive bytecode transformations reuse the same textual identifier for distinct entities (\eg, \texttt{a3} could be the name of both a class and a method) or exploit naming rules that are valid in bytecode but hard to include in decompiled Java source~\cite{DBLP:journals/jss/ChanY04}.
For example, a method could be named \texttt{if} or \texttt{a-b} in a valid class file~\cite{jvms26-unqualified-names}, but neither name can be expressed as a method identifier in compilable Java source.
When an obfuscated project can be compiled, such malicious renaming can often be countered by systematically assigning legal and unambiguous names directly in the bytecode, \eg, through alpha-renaming~\cite{Sabel2017, Beringer2022AlphaRenaming}.

Reverse engineers rarely have a complete codebase that compiles and is internally consistent.
We therefore generally restrict the available code to the method we want to deobfuscate and progressively larger amounts of surrounding context. 

\textbf{Identifier deobfuscation in Java.}
With limited opportunities to execute the obfuscated code, Java identifier recovery in a static setting is generally approached through signature or fingerprint matching, or structural inference with statistical name prediction~\cite{DBLP:conf/dais/BaumannPM17,DBLP:conf/ccs/BichselRTV16}. 
Existing approaches use a variety of techniques. 
Anti-ProGuard~\cite{DBLP:conf/dais/BaumannPM17} matches code against known unobfuscated applications using SimHash and $n$-gram fingerprints. 
DeGuard~\cite{DBLP:conf/ccs/BichselRTV16} jointly predicts identifiers from an application dependency graph, while random-forest methods recover method-name verbs from opcode features~\cite{DBLP:conf/snpd/IsobeT18}.  
Finally, DOBF~\cite{DBLP:conf/nips/LachauxRSL21} uses a transformer to reconstruct identifiers from source context.

\subsection{Large Language Models}\label{ssec:llm}
We provide a brief overview of LLMs, their general characteristics and the contemporary releases and orchestration techniques that are relevant to our study. 

\textbf{Generalities.}
LLMs are a class of machine learning models that process and generate language and information.
Most LLMs are based on the Transformer architecture~\cite{DBLP:conf/nips/VaswaniSPUJGKP17} to focus on the relevant parts of a text when generating the output. 
Trained on large corpora of text from different sources, they capture statistical patterns and structures of language.
LLMs typically follow a process of pre-training on generic tasks, sometimes fine-tuned or adapted afterwards on specific tasks and corpora.
Thanks to their complexity and size, these models are capable of generalizing learned knowledge, capable of applying patterns and information from the training data to new, unseen situations, and solving a wide range of tasks.
In this paper, we evaluated the effectiveness of three LLMs for the deobfuscation task (see \autoref{ssec:models-and-metrics}).

\textbf{Contemporary releases.}
LLM development has seen many breakthroughs in autoregressive pre-training, data curation, code-specific training, and mixture-of-experts (MoE) architectures.
Recently, reasoning-oriented post-training and knowledge distillation surfaced: smaller models can inherit capabilities from flagship models through knowledge distillation (\eg, a smaller \emph{student} model is trained on outputs, reasoning traces or probability distributions produced by a larger \emph{teacher} model).
A summary of these technical advances can be found in \citet{DBLP:journals/corr/abs-2303-18223}.
Qwen and DeepSeek have become major competitors in the frontier AI race, particularly for publicly available reasoning and code models (\ie, models designed for step-by-step problem solving and programming).
Our selected models (see \autoref{ssec:models-and-metrics}) represent several branches of this frontier evolution.
Qwen3-30B-A3B~\cite{qwen3}, for instance, uses such strong-to-weak distillation, is a MoE model, and supports a reasoning mode. 
We disable this mode in the main evaluation for comparability with the other models, as explained in \autoref{ssec:models-and-metrics}.

\textbf{RAG.}
Retrieval-Augmented Generation (RAG) is a technique that combines the generative capabilities of LLMs with external knowledge sources, such as databases or document collections, to improve the accuracy and relevance of generated responses~\cite{DBLP:conf/nips/LewisPPPKGKLYR020}.
It involves building an index of relevant documents, retrieving the highest ranking ones based on the input query using a carefully chosen retrieval score, and then feeding the LLM with this retrieved information.

\textbf{Agentic LLMs.}
A contemporary trend in LLM orchestration is to use agentic LLMs: there, an agent uses an LLM as a reasoning and decision-making component within an iterative perception–action loop. 
Given a goal, an agent can analyze the current context and plan coming steps by selecting actions such as retrieving information, calling external tools that \eg execute code \etc 
The results of these actions and decisions are returned to the agent as observations, which it can use to update its strategy and decide what steps to take next.
The agent is endowed with a memory mechanism that preserves relevant information across steps and reflection units can detect errors and revise previous decisions.
This combination of planning, tool use, memory and feedback looping enables the agent to perform multi-step tasks with greater autonomy than a standard LLM that generates a single response~\cite{DBLP:journals/fcsc/WangMFZYZCTCLZWW24}.

\textbf{LLM-based deobfuscators.}
LLMs themselves can be used for deobfuscation.
Besides VarBERT~\cite{DBLP:conf/sp/PalBBDNBGSASDBW24} and GENNM~\cite{DBLP:conf/ndss/Xu0SH0Y0X00025}, we acknowledge \textsc{SYMGEN}, a deobfuscator that targets only function names. 
It masks decompiler-generated function identifiers and generates replacements with an LLM adapted to decompiled code by training on a function-summarization task~\cite{DBLP:conf/ndss/JiangJ025}. 
All the aforementioned LLM approaches recover symbols lost as a consequence of native compilation and stripping.
In our study, instead, we target class, method, and field identifiers deliberately renamed in Java bytecode.
This matters because meaningful names and deliberately obfuscated names can coexist in Java.
Moreover, we study recovery under increasingly rich program and repository contexts, a problem typically left aside from previous work.

\section{Methodology}
\label{sec:design}
We lay out the methodology of our identifier-recovery study, including the considered context levels, the models taken into account, and the metrics we use to measure performance.

\subsection{Task definition}
\label{ssec:task-definition}

We consider a Java repository in which class, method, and field names have been consistently replaced (\ie obfuscated) with meaningless identifiers.
Identifiers belonging to external libraries and frameworks are not recovery targets and remain clear when they would normally survive obfuscation, as is the case for the three obfuscators considered in this work.
Each benchmark instance focuses exclusively on identifiers occurring inside a \emph{focal-method} $m$, \ie, the method of interest we try to deobfuscate.
Depending on the evaluation settings (see \autoref{ssec:settings}), the model under test receives some obfuscated repository context, $C(m)$, whose additional identifiers are not prediction targets.
That is, the repository we draw $C(m)$ from is the one $m$ originates from.
We use $m$ as an evaluation unit because it captures a coherent piece of program logic while keeping the input manageable, as in other attempts~\cite{DBLP:conf/sp/PalBBDNBGSASDBW24}.
Focal-methods are drawn entirely from project-owned code, \ie, code compiled from the main source (\eg \texttt{src/main/java}) and not belonging to a Java standard library or a third-party dependency.
We additionally exclude tests and code located under \texttt{third\_party/}, \texttt{examples/}, \texttt{vendor/} or \texttt{demo/} directories.

For a given obfuscator, let \(M^{\mathrm{clear}}\) and \(M^{\mathrm{obf}}\) denote the sets of clear and obfuscated focal-methods produced by the collection and curation pipeline we describe in \autoref{ssec:data-collection}.
Each retained pair \((m^{\mathrm{clear}},m^{\mathrm{obf}})\) contains the clear and obfuscated versions of the same method.
We tap into the obfuscation process to obtain original-to-obfuscated private mappings and to match methods using the declaring class and JVM method descriptor as well as the private mapping.
Each mapped class, method or field declaration has a unique private mapping ID linking its clear and obfuscated versions, even when distinct declarations share the same name.
Only uniquely resolved correspondences are retained.
Moreover, absent (or otherwise ambiguous) matches are discarded.
Thus, for each obfuscator, \(M^{\mathrm{clear}}\) and \(M^{\mathrm{obf}}\) are in bijection, although the resulting populations may differ across obfuscators by design (\eg, yGuard generates a different number of obfuscated methods compared to Allatori).
The $m^{\mathrm{clear}}$ methods are never exposed to the models or retrievers during inference.
They are retained as references and for contamination auditing; scoring uses the original names and target identities recorded in the private mappings, as detailed below.

\begin{figure}[!t]
  \centering
  \includegraphics[width=\textwidth]{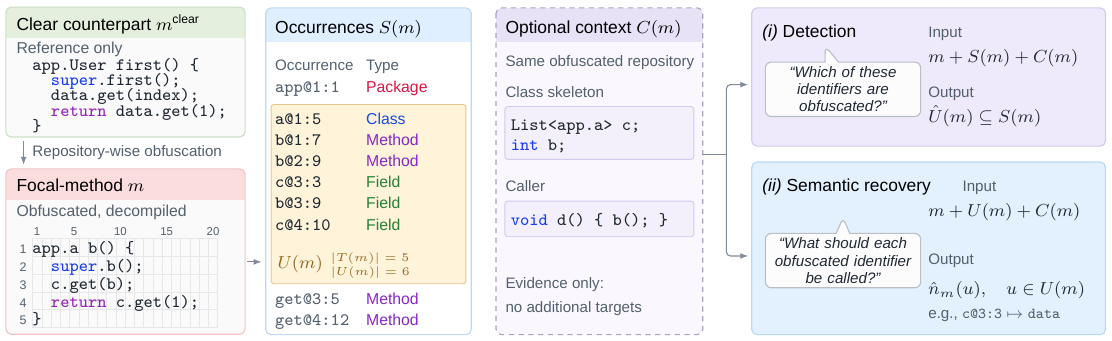}
  \vspace{-0.5em}
  \caption{Inputs and outputs of the identifier-detection and semantic-recovery tasks.
{The highlighted subset $U(m)$ contains six occurrences of five distinct targets in $T(m)$.}
{Semantic recovery receives $U(m)$.}}
  \label{fig:detection-semantic-recovery}
\end{figure}

\autoref{fig:detection-semantic-recovery} illustrates the distinction between target occurrences and the program declarations to which they refer.
For any obfuscated focal-method \(m \in M^{\mathrm{obf}}\), we use the notation in \autoref{tab:identifier-notation}.

\begin{table}[t]
  \centering
  \caption{Identifier occurrences and distinct targets in a focal-method $m$.}
  \label{tab:identifier-notation}
  \small
  {
\begin{tabularx}{\linewidth}{@{}lX@{}}
    \toprule
    \textbf{Notation} & \textbf{Meaning} \\
    \midrule
    $S(m)$ & All visible identifier occurrences in $m$. \\
    $U(m)\subseteq S(m)$ & Identifier occurrences whose project-owned class, method, or field declaration was genuinely obfuscated. \\
    $T(m)$ & The different declarations referred to by $U(m)$; private mapping IDs distinguish them. \\
    \bottomrule
  \end{tabularx}
}
\end{table}

The public candidate inventory $S(m)$ is extracted from the method body and identifies target occurrences by their position (\eg, \texttt{a@line:column}).
It may also include unchanged identifiers, such as library and package names, which are absent from $U(m)$.
Target occurrences may include the focal-method's own name, invoked methods, accessed fields, and classes used as types, constructors, casts, or static receivers.
The private mapping defines a map:
{
\[
  \pi_m:U(m)\longrightarrow T(m),
  \qquad T(m)=\{\pi_m(u):u\in U(m)\}.
\]
}
Several occurrences can refer to one target, which means the cardinalities satisfy $|T(m)|\leq |U(m)|$.
Conversely, identical spellings don't necessarily refer to the same declaration in Java.
For example, the occurrences of \texttt{a} in \texttt{server.struct.a a = server.struct.a.getInstance()} need not share a declaration or original name.
Such collisions occur in our datasets obtained with ProGuard, yGuard, and Allatori.
Distinct targets can also share the same clear name, but their private mapping IDs remain different.

In \autoref{fig:detection-semantic-recovery}, the two occurrences \texttt{c@3:3} and \texttt{c@4:10} are linked to the same target.
Both refer to the field originally named \texttt{data}.
By contrast, \texttt{b@1:7} and \texttt{b@3:9} refer to two different targets: the method originally named \texttt{first} and the field originally named \texttt{index}.
Their names became identical only after obfuscation.
There is also a third case: the declaration \texttt{b@1:7} and the call \texttt{b@2:9} in \texttt{super.b()} refer to different methods, even though both were already named \texttt{first} before obfuscation.
One is declared in the focal-method's class, while the other is the superclass implementation: this is a case where distinct code elements bear the same clear name and the same obfuscated name, but they are still distinguishable by their private mapping IDs.

This third case occurs in our dataset: one ProGuard focal-method declares \texttt{i()} and calls \texttt{super.i()} inside its body.
Both methods were originally named \texttt{registerGoals}, but the first belongs to \texttt{SoldierZombie} and the second to \texttt{PazZombie}.
They have separate private mapping IDs despite sharing the same name in both versions of the code.
From this example, the call \texttt{super.i()} tells us that a superclass method is invoked; however, identifying \texttt{PazZombie} as its declaring class requires the class structure and hierarchy residing outside the focal-method.
Because of all these corner cases, assessing the quality of the models' outputs throughout this paper will rely on the repository-level identities and the private mapping created during obfuscation instead of on the sole code body of $m^{\text{clear}}$: matching names between the clear and obfuscated methods would not distinguish these targets by itself.

Private mapping IDs and the map $\pi_m$ are never exposed to the model.
Detection addresses the occurrences within the inventory it is provided, $S(m)$.
Semantic recovery is done in a semi-oracle setting as the model receives the target-occurrence inventory $U(m)$; the underlying target identities in $T(m)$ remain private.
The underlying $\pi_m$ is thus transparent to the evaluated model, but the distinction between $U$ and $T$ matters for the scoring; it uses private mapping IDs to determine which target each target occurrence refers to. 
Repeated occurrences of the same target count once; different targets are scored separately, even when they share the same spelling.

Given a focal-method \(m\), we distinguish two related tasks: \textit{(i)} detecting which identifiers, visible inside the syntax of the entire method, have been obfuscated, and \textit{(ii)} recovering meaningful names for known obfuscated identifiers.
This separation reflects common reverse-engineering practice, where analysts progressively identify program elements of interest and assign them semantic roles as their understanding of the program develops~\cite{DBLP:conf/uss/VotipkaRMFM20}.
This practice is also consistent with \textsc{DeGuard}, which preserves identifiers that remain meaningful while inferring names for identifiers that have been obfuscated~\cite{DBLP:conf/ccs/BichselRTV16}.
A deobfuscator can miss an obfuscated identifier altogether, or detect it and then choose a name with the wrong spelling or meaning. 
These related tasks have different failure modes, so we evaluate them separately in the main experiments and combine them in the ablation study.

In the \emph{detection} task \textit{(i)}, the model receives the candidate inventory \(S(m)\) and predicts a subset
{
\[
    \widehat{U}(m) \subseteq S(m)
\]
}
containing the occurrences that it considers obfuscated.
In the \emph{semantic recovery} task \textit{(ii)}, the model receives the ground-truth occurrence inventory $U(m)$ and predicts a replacement name $\widehat{n}_m(u)$ for each occurrence $u\in U(m)$.
Different occurrences of the same target may receive different predictions, within one focal-method or across several methods. 
We detail in \autoref{ssec:models-and-metrics} that the error is computed at the occurrence level against the target's true name but averaged over each focal-method.
The reference name for occurrence $u$ is $n(\pi_m(u))$.
It is the required output for exact match, although another name can obtain a high similarity score under metrics we describe further down.
The prediction itself is required to be a valid Java identifier.
However, we do not require the program obtained by substituting the predicted names to compile because, during the semantic recovery task, our primary objective is to assess the quality of name recovery: this notably includes results where the same lexeme is assigned different names if they refer to different targets (\eg, one occurrence of \texttt{a} can be assigned \texttt{balance} and the other \texttt{timeout}).

Given a fixed focal-method $m$ and its ground-truth target set $T(m)$, what the model receives depends on the task and experimental setting:
\begin{itemize}
    \item For detection \textit{(i)}, the model receives at least the focal-method and the public candidate inventory \(S(m)\).
    \item For semantic recovery \textit{(ii)}, the model receives at least the focal-method and the real target-occurrence inventory \(U(m)\) with the corresponding type information.
    \item Depending on the context setting described in \autoref{ssec:settings}, either task may additionally receive code from the same obfuscated repository as \(m\) was drawn from. We call this additional context $C^{\text{setting}}(m)$, with \quotes{setting} being one of $L_0$, $L_1$, $L_2$, $L_3$, $R$, or $A$, as detailed in \autoref{ssec:settings}. For $L_0$, the additional context is empty: $C^{L_0}(m)=\emptyset$.
    We denote it $C(m)$ for brevity, but it should be understood as being conditioned on the setting during the experiments.
    Note that a setting determines how much contextual evidence is allowed for the model to see, but the obfuscated identifiers within it are not themselves added to the target set of the current benchmark instance.
\end{itemize}

\subsection{Traditional identifier-recovery approaches as baselines}
Our work assesses the effectiveness of some traditional methods for deobfuscating our lexemes of interest.
\autoref{tab:deobfuscator_scope} summarizes the scope of these traditional deobfuscation systems and identifies those used in our study.
We exclude JADX~\cite{jadx} from the evaluated baselines because its detection capabilities are solely length-based and it assigns generic aliases without attempting to recover meaningful names; in this respect, it acts more like a lexeme formatter than a semantic recovery tool.
For each used tool, \ie Isobe-Tamada~\cite{DBLP:conf/snpd/IsobeT18}, Anti-ProGuard~\cite{DBLP:conf/dais/BaumannPM17}, DeGuard~\cite{DBLP:conf/ccs/BichselRTV16} and DOBF~\cite{DBLP:conf/nips/LachauxRSL21}, \autoref{ssec:traditional-baselines} describes how we used or reimplemented it.
Similarly to the LLM benchmarking, we first evaluate the ability of each one of the considered techniques to detect obfuscated identifiers.
Then, we measure their ability to recover the original names.
The main difference is that the traditional baselines do not natively separate tasks \textit{(i)} detection and \textit{(ii)} semantic recovery.
Instead, they do it in one single run, so the task differentiation is made only after the deobfuscated snippet is output, during the analysis phase.
We note that this apparently gives an upper hand to the LLMs for the semantic recovery task since they are given the ground-truth target-occurrence inventory $U(m)$ during the main evaluation campaign. 
The end-to-end ablation measures how recovery scores decrease when model-detected occurrences replace the oracle inventory (see \autoref{ssec:end-to-end-recovery}).
Overall, we will find that some methods reliably flag identifier obfuscation for the detection task, and retrieval occasionally generates a useful name for the semantic recovery task.
However, none of them nearly comes close to reconstructing the developer's vocabulary at a rate that would support program understanding.
Our four baselines average only $1.86\%$ exact-match recovery over methods with predictions, as we will see in \autoref{ssec:semantic-recovery-results}.
By comparison, VarBERT's user study selected functions with at least $75\%$ correctly recovered variable names and found that code-understanding scores increased from $55.49\%$ to $66.67\%$~\cite{DBLP:conf/sp/PalBBDNBGSASDBW24}.

\begin{table}[t]
    \color{black}
    \centering
    \caption{Scope and availability of traditional identifier recovery systems. F, M, and C denote field-, method-, and class-name recovery targets.}
    \label{tab:deobfuscator_scope}
    \footnotesize
    \setlength{\tabcolsep}{3pt}
    \renewcommand{\arraystretch}{1.08}

    \begin{tabular*}{\textwidth}{
        @{\extracolsep{\fill}}
        p{5.2cm}
        c
        cc
        c
        c
        @{}
    }
        \toprule
        \textbf{Deobfuscator}
        & \shortstack{\textbf{Scope}}
        & \shortstack{\textbf{Open}\\\textbf{source}}
        & \shortstack{\textbf{Signature-}\\\textbf{based}}
        & \shortstack{\textbf{Needs full}\\\textbf{project}}
        & \textbf{Used} \\
        \midrule

        JEB~\cite{jeb}
        & FMC$^{\mathrlap{\dagger}}$
        & --
        & \textcolor[HTML]{4287F5}{\(\checkmark\)}
        & --
        & -- \\

        JADX~\cite{jadx}
        & FMC$^{\mathrlap{\dagger}}$
        & \textcolor[HTML]{4287F5}{\(\checkmark\)}
        & --
        & --
        & -- \\

        Isobe--Tamada~\cite{DBLP:conf/snpd/IsobeT18}
        & M$^{\mathrlap{\dagger}}$
        & --
        & --
        & --
        & \(\textcolor[HTML]{4287F5}{\checkmark}^{\mathrlap{*}}\) \\

        Anti-ProGuard~\cite{DBLP:conf/dais/BaumannPM17}
        & MC
        & \textcolor[HTML]{4287F5}{\(\checkmark\)}
        & \textcolor[HTML]{4287F5}{\(\checkmark\)}
        & \textcolor[HTML]{4287F5}{\(\checkmark\)}
        & \(\textcolor[HTML]{4287F5}{\checkmark}^{\mathrlap{*}}\) \\

        DeGuard~\cite{DBLP:conf/ccs/BichselRTV16}
        & FMC
        & --
        & --
        & \textcolor[HTML]{4287F5}{\(\checkmark\)}
        & \(\textcolor[HTML]{4287F5}{\checkmark}^{\mathrlap{*}}\) \\

        DeProDexcator~\cite{deproDexcator}
        & FMC$^{\mathrlap{\dagger}}$
        & --
        & --
        & \textcolor[HTML]{4287F5}{\(\checkmark\)}
        & -- \\

        DOBF~\cite{DBLP:conf/nips/LachauxRSL21}
        & FMC
        & \textcolor[HTML]{4287F5}{\(\checkmark\)}
        & --
        & --
        & \(\textcolor[HTML]{4287F5}{\checkmark}^{\mathrlap{*}}\) \\

        \bottomrule
    \end{tabular*}

    \vspace{3pt}
    \begin{minipage}{\textwidth}
        \scriptsize
        \textcolor[HTML]{4287F5}{\(\checkmark\)}: yes;
        --: no;
        \(\cdot^{\dagger}\) indicates identifier handling without semantic name recovery.
        \textcolor[HTML]{4287F5}{\(\checkmark\)}$^{*}$ indicates that our implementation is an adaptation rather than the original system.
    \end{minipage}
\end{table}

\subsection{Context and Retrieval Settings} 
\label{ssec:settings}

We separate the experiments into six settings: static context scopes ($L_0$, $L_1$, $L_2$, $L_3$), repository-retrieval mechanism ($R$), and agentic context-retrieval ($A$).
This separation provides a way to assess the effects of context content, retrieval strategy and decision-making process.

\textbf{Static context scope.}
The $L$ setting controls which program evidence is supplied without repository-wise retrieval in a deterministic manner.
\begin{figure}[t]
  \centering
    \vspace{-0.5em}
    \includegraphics[width=0.75\linewidth]{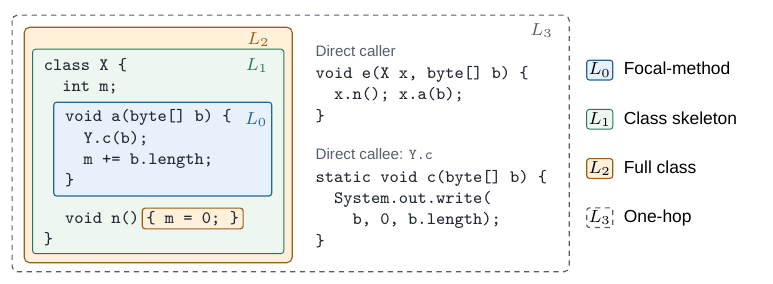}%
  \caption{Static context levels for identifier recovery.}
  \label{fig:static-context-levels}
\end{figure}
The four amounts of code that the LLM is provided with inside $C^{L_i}(m)$, depicted by example snippets inside Figure~\ref{fig:static-context-levels}, follow a nested relationship: $L_0 \subset L_1 \subset L_2 \subset L_3$.
\begin{itemize}
\item $L_0$: the model only has access to the focal-method's body and signature. This means $C^{L_0}(m)=\emptyset$. Project-owned class, method, and field names are rendered opaque as per the obfuscator's default setting.
Here we test recovery from the immediately available local code.
\item $L_1$: we supply a skeleton of the enclosing class. It contains the obfuscated class declaration and its modifiers, superclass, implemented interfaces, and fields, plus constructor and same-level-method signatures. The bodies of same-level methods are hidden. This shallow view still shows the structure of objects and how their types relate.
\item $L_2$: we show the entire enclosing class, including the bodies of same-level methods. The model can see usage across methods in the class, such as initialization of a field in one method and consumption in another. 
\item $L_3$: we augment $L_2$ with one-hop caller and callee method bodies when they are available (nothing is added otherwise).
Previous identifier-recovery work shows that caller and callee information can provide contextual evidence beyond what an isolated function can offer~\cite{DBLP:conf/ndss/Xu0SH0Y0X00025}.
We construct this context using static call-graph relations, with availability and composition summarized in \autoref{app:l3-caller-callee-evidence} and \autoref{tab:l3_caller_callee_availability}.
\end{itemize}

\textit{Motivation.}
We increase context in controlled steps, using work units drawn from reverse-engineering practice.
An analyst examining a routine found through a crash or execution trace~\cite{DBLP:conf/kbse/MuGCCGXMS19} has the focused view represented by the focal-method-only setting ($L_0$).
We introduce the class skeleton ($L_1$) mainly to control for the contribution of structural context. Inspecting a chosen decompiled class, or a partially recovered artifact, gives the complete enclosing class ($L_2$). If direct callers and callees ($L_3$) are also examined, interactions with those neighboring routines may help the analyst understand the focal-method.
These scopes are factored into broader code-inspection strategies featured in reverse-engineering practices~\cite{DBLP:conf/uss/VotipkaRMFM20}.
When the complete obfuscated repository is available, $R$ and $A$ instead test two ways to select supporting evidence: fixed relevance-based retrieval and model-selected inspections of ambiguous occurrences.
In $A$, inspections are selected together and executed in one round; the setting does not involve iterative navigation based on previously returned evidence.

\textbf{Repository-retrieval mechanism.}
$R$ is a setting where we use retrieval-augmented generation over the obfuscated repository only. 
It follows repository code completion approaches that retrieve evidence from elsewhere in the codebase, as in RepoCoder~\cite{DBLP:conf/emnlp/ZhangCZKLZMLC23}.
It relies on a conceptually simple key-value retrieval system that combines \textit{sparse} and \textit{dense} retrieval, as used in ReACC~\cite{DBLP:conf/acl/LuDHGHS22}, but requires the repository to undergo a preprocessing step.
\begin{enumerate}
    \item First, it is split into typed retrieval chunks: method bodies, class skeletons, complete short classes, field-access snippets and caller/callee snippets when available.
    We keep a top-level class whole if it contains at most 120 tokens, counted using UniXcoder's tokenizer. Field-access snippets collect each access line and the lines immediately before and after it, staying within the same method.
    Resources and configuration files are discarded to avoid leaking semantic names that may survive obfuscation.
    \item Second, Tree-sitter identifies identifier roles, which are encoded as retrieval terms (\eg, \texttt{identifier:method:b}). The stored source code remains unchanged, and these role terms are not shown to the LLM.
    We take this step because initial trials showed that the model benefited from the chunks recovered in that form.
    These role terms and the code's syntax features form the keys of the \textit{sparse} retrieval.
    \item Third, UniXcoder, a pretrained neural model~\cite{DBLP:conf/acl/GuoLDW0022}, encodes every annotated repository chunk into vectors which act as keys for the retrieval: they are stored in FAISS, a library that indexes numerical vectors and enables similarity retrieval~\cite{DBLP:journals/tbd/DouzeGDJSMLHJ26}.
    In this state, the embedded vectors represent the keys of the \textit{dense} retrieval system.
\end{enumerate}
The focal-method $m$ queries both retrievers after its own chunk has been removed from the candidate pool.
On one side, BM25 ranks the indexed repository chunks by their token overlap with the query; it gives more weight to unusual tokens and takes into account token frequency~\cite{DBLP:journals/ftir/RobertsonZ09}.
This produces a ranked list $L_{\mathrm{bm25}}(m)$ of up to 100 chunks.
On the other side, UniXcoder encodes the focal-method into a vector which is compared with the vectorized chunks stored in FAISS to find the closest ones in contextual similarity, \ie a ranked list $L_{\mathrm{dense}}(m)$ of the top-100 chunks.
For long focal-methods, we encode overlapping excerpts within UniXcoder's 512-token input limit (including special tokens), then average and renormalize their vectors into a single query vector, using the chunk-and-average approach also evaluated in LongEmbed~\cite{ZhuEtAl2024LongEmbed}.
To combine the lists, we use Reciprocal Rank Fusion (RRF): \textit{rank-level fusion} is standard in information retrieval and repository-level code generation~\cite{DBLP:conf/sigir/CormackCB09,PanEtAl2025CatCoder}, and avoids comparing raw BM25 and embedding-similarity scores measured on different scales.
Ranks start above zero. For either top-100 list, a missing chunk has no contribution. We use $60$, the original RRF paper's constant for damping the influence of the highest ranks. We include chunks in $s_R(m,c)$ order until the fixed 8k-token context budget is reached. For candidate chunk $c$ and query method $m$, the score is
\[
s_R(m,c)
=
\frac{\mathbf{1}[c \in L_{\mathrm{bm25}}(m)]}
     {60 + \operatorname{rank}_{\mathrm{bm25}}(m,c)}
+
\frac{\mathbf{1}[c \in L_{\mathrm{dense}}(m)]}
     {60 + \operatorname{rank}_{\mathrm{dense}}(m,c)} .
\]

\textbf{Agentic retrieval.}
We design a simple task-specific harness for the $A$ setting; it uses iterative, model-controlled retrieval.
The repository-navigation tools are inspired by SWE-agent~\cite{DBLP:conf/nips/YangJWLYNP24}. To choose evidence for a sample, the model asks for structural information about a selected identifier in the obfuscated repository: its definition, enclosing class, callers, callees, or field uses. We give it read-only tools, without clear source or obfuscation mappings, file modification, or arbitrary command execution. Each sample has one retrieval round before its final prediction: up to two inspections can return a combined 700 evidence tokens. We save the entire interaction before private scoring of the predictions, including tool calls, the evidence returned, and token usage.

\subsection{Models and Metrics}
\label{ssec:models-and-metrics}

We select models representative of the state of the art in open-weight code LLMs as of August 2026, with sizes suitable for practical academic use.
\begin{itemize}
    \item \textbf{Qwen3-30B-A3B} is a general-purpose mixture-of-experts (MoE) model with \(30.5\) billion parameters, approximately \(3.3\) billion of which are activated per token.
    It supports both reasoning and direct-response modes.
    We disable reasoning in the main evaluation to compare the models in a common direct-response setting, since the selected OLMo Instruct checkpoint does not offer an equivalent reasoning mode~\cite{olmo31ModelCard}. 
    It was released on 29 April 2025~\cite{qwen3Release}.
    There is no officially disclosed knowledge cutoff for this model.
    Our runs used revision \texttt{ad44e777} in direct-response mode. Decoding was deterministic: temperature 0, top-$p=1$, and seed 20260818. The model context was 40,960 tokens, with a limit of 4,096 output tokens. 
    We refer to this model as Qwen later on.
    \item \textbf{{OLMo~3.1-32B-Instruct}} targets language understanding and coding.
    We use \texttt{allenai/Olmo-3.1-32B-Instruct}, revision \texttt{ac0587e4}, and refer to it as OLMo throughout the results.
    Its training-data cutoff is December~2024. 
    Unlike models that release only their weights, OLMo is an unusual case in the contemporary LLM landscape as the authors provide the complete model-development flow, including training data, code, recipes, intermediate checkpoints, and data provenance, which enables end-to-end auditability and reproducibility~\cite{olmo3}.
    We can therefore check for training-data contamination, a check that is generally unavailable when a model's training corpus is undisclosed.

    \item \textbf{DeepSeek-V4-Flash-0731} is an open-weight frontier MoE model targeting coding and agentic tasks. 
    It was released on 31 July 2026~\cite{deepseekv4flash0731} but the company does not establish its training-data cutoff.
    Despite activating only 13B parameters per token, it outperforms DeepSeek-V4-Pro (Preview) and GLM-5.2 across the shared benchmarks reported by DeepSeek, and performs on par with Claude Opus~4.8 on Terminal-Bench~2.1 (82.7 vs.\ 85.0) and Agents' Last Exam (25.2 vs.\ 25.7)~\cite{deepseekv4flash0731}.
    We call it DeepSeek for the rest of the study.
\end{itemize}

\textbf{Definition of identifier types.}
{Each occurrence in $S(m)$ has a scoring label given by a map}
{\(\vartheta_m:S(m)\rightarrow\{\mathrm{F},\mathrm{M},\mathrm{C},\mathrm{Other}\}\),}
{where F, M, and C denote field, method, and class identifiers, and Other denotes other or unresolved roles.}
{We obtain these labels with a Java parser~\cite{TreeSitterJava} and repository-wide bytecode analysis using SootUp~\cite{KarakayaEtAl2024SootUp}, including declaration owners, call relationships, and type descriptors.}
{Occurrences whose roles remain ambiguous are assigned to Other, which accounts for 2\% of the total occurrences in our datasets.}
{These scoring labels are deliberately withheld from the input at the detection task; we do supply the types alongside the inventory $U(m)$ for the semantic recovery task.}
{We request inferred types during detection but do not evaluate type classification itself: the inferred types are used for our ablation study evaluating end-to-end identifier recovery (see Section~\ref{ssec:end-to-end-recovery}).}
{For each $k\in\{\mathrm{F},\mathrm{M},\mathrm{C}\}$, we define}
{
\[
 \begin{gathered}
 S_k(m)=\{s\in S(m):\vartheta_m(s)=k\},\\[3pt]
 U_k(m)=U(m)\cap S_k(m),\qquad T_k(m)=\pi_m(U_k(m)).
 \end{gathered}
\]
}
{Thus $U(m)=\bigsqcup_k U_k(m)$ and $T(m)=\bigsqcup_k T_k(m)$, using the special union symbol, represent disjoint partitions of target occurrences and target declarations, respectively.}
{This also means that packages, local variables, parameters, external-library identifiers, and unchanged names are not target occurrences in $U(m)$.}
{The detector predicts occurrences $\widehat{U}(m)\subseteq S(m)$; on the other hand, semantic recovery predicts names for the supplied occurrences in $U(m)$.}

\textbf{Metrics used for the detection task.}
{Predictions are done occurrence-wise, but detection is scored over distinct code elements within each focal-method to avoid over-counting (see \autoref{ssec:task-definition}).}
{We employ a \quotes{touch-any} rule: a target is considered detected if at least one of its occurrences is flagged.}
{The detected target set is}
{
\[
 \widehat{T}(m)=\{\pi_m(u):u\in\widehat{U}(m)\cap U(m)\}\subseteq T(m).
\]
}
{This set is derived only privately for scoring; it is not what the detector outputs, \ie, $\widehat{U}(m)$.}
{The true-positive and false-negative counts therefore are}
{
\[
 \mathrm{TP}_m=|\widehat{T}(m)|,\qquad
 \mathrm{FN}_m=|T(m)\setminus\widehat{T}(m)|.
\]
}
{Repeated detections of one target within a focal-method therefore give no additional credit, and missed occurrences of an already detected target add no false negatives.}
{Predictions on non-target identifiers increase $\mathrm{FP}_m$. Each declaration counts at most once per focal-method, whether its selection constitutes a true positive or a false positive.}
{Non-target identifiers disregarded by the detection are true negatives that do not enter $F_1$.}
We exclude true negatives because unobfuscated identifiers are often numerous and credit for leaving them unchanged can dilute the poor detection of obfuscated targets.
{For each focal-method, we compute}
{
\[
 \begin{gathered}
 F_{1,m}=\frac{2\mathrm{TP}_{m}}{2\mathrm{TP}_{m}+\mathrm{FP}_{m}+\mathrm{FN}_{m}}.
 \end{gathered}
\]
}
{Detection $F_1$ is then macro-averaged over focal-methods, so each method contributes equal weight to this score.}

{For scores restricted to type $k$ (we also refer to them as \textit{type-specific} scores as opposed to \textit{type-agnostic} scores), true positives and false negatives use the target's original type:}
{
\[
 \mathrm{TP}_{m,k}=|\widehat{T}(m)\cap T_k(m)|,\qquad
 \mathrm{FN}_{m,k}=|T_k(m)\setminus\widehat{T}(m)|.
\]
}
{Non-target flags are grouped by declaration and assigned to $\mathrm{FP}_{m,k}$ according to their scoring label $\vartheta_m(s)=k$, independently of the model's inferred type.}
{A correctly resolved target receives detection credit even when its predicted type is wrong.}
We compute
{
\[
 \begin{gathered}
 F_{1,m,k}=\frac{2\mathrm{TP}_{m,k}}{2\mathrm{TP}_{m,k}+\mathrm{FP}_{m,k}+\mathrm{FN}_{m,k}}.
 \end{gathered}
\]
}
For each method $m$ and identifier type $k$, we compute $F_{1,m,k}$ only when its denominator is nonzero. 
We then average these scores across all focal-methods to obtain $F_{1,k}$.
{The overall $F_1$ combines all types within each method before averaging (so it is not an average of $F_{1,\mathrm{F}}$, $F_{1,\mathrm{M}}$, and $F_{1,\mathrm{C}}$).}
{We emphasize that these type-restricted scores describe detection performance for fields, methods, and classes; they do not measure the accuracy of the model's inferred types: this faculty is not evaluated in this work.}

\textbf{Metrics used for the semantic recovery task.}
{For semantic recovery, the model receives a focal-method $m$ and the ground-truth occurrence inventory $U(m)$. 
Predicted names are compared with the original names of their private targets using lexical and semantic similarity metrics.}
Definitions of these lexical metrics are available in prior work~\cite{MuLim2023TextSimilarity,lin-2004-rouge,TervenEtAl2025Metrics}.
Exact Match (\texttt{ExM}) compares the original identifier spellings, including case. For the graded metrics (which give partial credit), we split identifiers at delimiters, camel-case and acronym transitions, and letter/digit boundaries, then case-fold the components. Longest common subsequence (LCS) and Levenshtein similarity (LevSim) compare these components concatenated without separators. LCS length is divided by the longer normalized string's length; LevSim is one minus the edit distance divided by that length. Jaccard (\texttt{Jac}) divides the size of the intersection of component sets by the size of their union, while ROUGE-L $F_1$ (RL-$F_1$) compares the ordered component sequences.

However, these metrics merely assess surface-form overlap and could assign low scores to otherwise semantically good predictions expressed using different words.
We therefore additionally assess \emph{semantic similarity} using BERTScore~\cite{DBLP:conf/iclr/ZhangKWWA20}.
Before computing BERTScore, we join the case-folded identifier components with spaces (\eg, \texttt{personName} becomes \texttt{person name}).
Identifier splitting is a well-established preprocessing operation for extracting the lexical components of source-code identifiers~\cite{DBLP:journals/ese/HillBLPV14}.
The same preprocessing and fixed encoder are applied to every prediction and ground truth.
The text is then divided into subtokens using the tokenizer of DeBERTa-xlarge-MNLI~\cite{DBLP:conf/iclr/HeLGC21}, a pre-trained Transformer encoder model.
This same model then assigns each subtoken an embedding vector reflecting its context within the split name (the surrounding Java code is not supplied to this encoder, only the identifiers' worth of subtokens).

BERTScore compares the obtained subtoken embeddings of a predicted identifier $\widehat{y}$ with the subtoken embeddings of its ground-truth identifier $y$ using the cosine similarity of these vectors.
For each predicted subtoken, it finds the reference subtoken with the most similar vector; averaging these maximum similarities gives semantic precision $P_{\mathrm{B}}$, which measures how well the prediction is supported by the reference.
In the notation below, predicted subtokens \(\widehat{t}_1,\ldots,\widehat{t}_p\) have contextual embeddings \(\widehat{\mathbf h}_i\), and reference subtokens \(t_1,\ldots,t_q\) have contextual embeddings \(\mathbf h_j\).
This gives \(P_{\mathrm{B}} = (1/p)\sum\nolimits_{i=1}^{p}\max\nolimits_{1\leq j\leq q}\cos(\widehat{\mathbf h}_i,\mathbf h_j)\).
Semantic recall reverses the comparison and averages over the reference subtokens: \(R_{\mathrm{B}} = (1/q)\sum\nolimits_{j=1}^{q}\max\nolimits_{1\leq i\leq p}\cos(\widehat{\mathbf h}_i,\mathbf h_j)\).
In this study, we report their harmonic mean, BERTScore $F_1$ (B$F_1$), computed as $2P_{\mathrm{B}}R_{\mathrm{B}}/(P_{\mathrm{B}}+R_{\mathrm{B}})$.
Unlike detection $F_1$, this score thus summarizes encoder-subtoken similarities, not counts of correctly and incorrectly flagged identifiers.
Using this metric, related words can receive credit even when their spellings differ.

{We additionally measure cosine similarity between UniXcoder representations~\cite{DBLP:conf/acl/GuoLDW0022} of the predicted and reference names, which we denote \textit{SemSim}.}
After the same identifier preprocessing, the predicted name $\widehat{y}$ and its ground-truth name $y$ are encoded separately.
SemSim is the cosine similarity between their final-layer CLS embedding vectors, where CLS is a special token at the start of each input. It compares one vector per name, whereas BERTScore compares individual subtoken embeddings.

{The model predicts names occurrence-wise, whereas scoring implies resolving those occurrences to the genuine targets that delivered them through $\pi_m$.}
{For each input occurrence, the prediction $\widehat{n}_m(u)$ is therefore associated with the ground-truth name $n(\pi_m(u))$.}
{Distinct targets with identical obfuscated lexemes (\eg, the two fields in \texttt{this.x + other.x;}) are therefore scored separately.}
{Several occurrences can refer to the same target and receive different predicted names; these predictions are evaluated separately against the ground-truth name and averaged so repeated predictions do not receive additional weight.}

{These method scores are then averaged with equal weight over the entire pool of focal-methods.}
{This \emph{method-macro} aggregation will count the same target again if it appears in another focal-method, but never weights repeated occurrences within one focal-method more heavily.}
We report subtoken ROUGE-L $F_1$ (sometimes abbreviated as RL-$F_1$)~\cite{lin-2004-rouge}, raw BERTScore $F_1$ (\textit{alias} B$F_1$) using DeBERTa-xlarge-MNLI, and SemSim.
Higher scores indicate closer agreement with the original names: RL-$F_1$ measures shared name components in order, B$F_1$ measures semantic similarity between subtoken embeddings, and SemSim compares embeddings of the complete names.

\section{Experimental Setup}
\label{sec:experimental-setup}

\subsection{Hardware, Data Collection and Contamination Control} 
\label{ssec:data-collection}
We list our used hardware and detail how we construct a benchmark of identifier-renaming deobfuscation from recently created open-source repositories under contamination control.

\textbf{Hardware.} 
We run the creation and preprocessing of the datasets, the training of the non-neural baselines, the processing of the JARs with the baselines and the scoring of the results on an Intel Xeon Phi 7230, 64 cores/256 hardware threads at 1.30 GHz with 125.6 GiB system RAM.
The inferences of Qwen and OLMo are run between two NVIDIA H100 NVL GPUs with 93.1 GiB of usable VRAM each, with 31 allocated Intel Xeon Platinum 8562Y+ vCPUs.
We queried DeepSeek directly through \url{https://api.deepseek.com}, requesting \texttt{deepseek-v4-flash}.
Our experiment logs identify it as \texttt{DeepSeek-V4-Flash-0731}.
The infrastructure supporting the service is unknown to us.

\textbf{Risks of contamination in LLM evaluation.} 
Since code LLMs can memorize public source code and code-benchmark contamination has been shown to improve apparent model performance, identifier deobfuscation requires particular care: a correct prediction may come from recognizing a previously seen code context and recalling the original name rather than from inferring a meaningful identifier from semantics~\cite{DBLP:conf/icse/Al-KaswanID24,DBLP:conf/acl/RiddellNC24}.
This concern is well documented in recent LLM evaluation work where benchmark contamination can inflate measured performance and obscure true generalization \cite{DBLP:conf/naacl/Deng0TGC24}. 
This is particularly true for code benchmarks, since public repositories and programming problems are common pretraining sources~\cite{DBLP:conf/ijcai/NiuL0022}; recent benchmarks therefore either rely on real but carefully dated software tasks \cite{DBLP:conf/iclr/JimenezYWYPPN24}, continuously updated or post-cutoff tasks \cite{DBLP:conf/iclr/WhiteDRPF0SJSDS25,DBLP:conf/iclr/JainHGLYZWSSS25}, or explicit contamination measurements for code-generation benchmarks \cite{DBLP:conf/acl/RiddellNC24}.
 
\textbf{Dataset Construction.} We seek to limit overlap between our dataset and the local models' likely training data. We want the models to infer names from the program evidence, with as little opportunity as possible to recall names seen during training.
We select recently created repositories and record the exact commit used for each project; their creation dates postdate Qwen and OLMo but not necessarily DeepSeek.
We first collect and locally audit 867 Java repositories from GitHub, spanning 1 August 2025 to 13 June 2026.
Our clean-source check consists of inspecting the build files and repository metadata. ProGuard, R8, minification settings, mapping files and other obvious obfuscation-related configurations are cues that the source may already be obfuscated.
We also check provenance, rejecting Java history that predates the cutoff, migration/fork-like wording, multiple Git roots and git submodules.
We attempt to build each audited commit. We discard a project if it fails to compile or produces no project-owned main class files. From its paths and build outputs, our heuristics retain the main Java code owned by the project. They exclude tests, examples, generated files, benchmarks, and artifacts that are not project-specific.
For the main selection, we require Maven or Gradle and enough owned Java code ($\geq 5{,}000$ LOC). There are 60 projects that compile before curation. Obfuscation, CFR decompilation and leakage checks leave 46. To supplement this set with large projects, we audit another 17 repositories: 12 compile and enter curation, and 5 pass the same bytecode and leakage checks.
Finally, we keep 51 repositories for the clear dataset.
For the repositories that pass these filters, we manually inspect the selected repositories for obvious identifier obfuscation such as meaningless names or generated generic identifiers.
We then compile the projects normally and collect the compiled \texttt{.class} files that appear to belong to project-owned code, again excluding test, generated, example, demo outputs.
Those selected class files are packaged into a JAR, obfuscated with ProGuard, yGuard and Allatori (we address these later in the section), and then decompiled with CFR~0.152.

\begin{table}[t]
\centering
\caption{Target counts by obfuscator, with counts of focal-methods inside the retained evaluation cohort of all three evaluated LLMs.
F, M, and C mean field, method, and class.}
\label{tab:identifier_target_stats}
\begin{tabular}{lccccc}
\toprule
Obfuscator & Focal-methods & F & M & C & Total \\
\midrule
Allatori & 58,065 & 71,508 & 118,212 & 73,462 & 263,182 \\
ProGuard & 58,721 & 74,855 & 126,875 & 70,823 & 272,553 \\
yGuard   & 59,078 & 79,233 & 137,764 & 62,543 & 279,540 \\
\bottomrule
\end{tabular}
\end{table}

To control for possible OLMo training-data contamination (it is the only model that offers this opportunity among the evaluated models), we compared exact file and syntax-normalized method hashes from the clear repositories against the disclosed datasets and corpora used for its training and pre-training.
They comprise the StackEdu corpus, a curated source-code dataset derived from The Stack~v2 and included in OLMo~3's training mixture~\cite{stackedu,dolma3,olmo3}.
They also encompass Dolci SFT, DPO, and RL data (including prompts and responses) and the code parts of Dolmino and Longmino~\cite{dolciSFT,dolciDPO,dolciRL}.
We considered an exact declaration match conclusive when it contained at least 50 tokens and appeared in fewer than 100 StackEdu documents; this rule avoids treating tiny or common methods (\eg, a standard getter method) as proof of contamination: 50 tokens would require enough distinctive code, while the 100-document limit aims at removing boilerplate code used in many different projects.
This audit found no exact full-file match but several method matches in the DPO data and the StackEdu collection.
Rather than removing entire repositories because they contained matching code, we excluded only focal-methods that directly satisfied this criterion: 249 Allatori methods, 188 ProGuard methods, and 236 yGuard methods were removed and all 51 repositories were retained.
The same exclusion list was applied consistently to the $L_0$-$L_3$ evaluation cohorts: that way, the removed methods cannot occur as either focal-method (\ie, $m$) or additional context (also denoted $C(m)$ in this work) in any of these conditions.

From the decompiled obfuscated code, we extract focal-methods, target identifiers, and the corresponding context levels used in the benchmark.
Focal-methods are normalized (spaces, tabs, comments and docstrings are stripped, extra line returns removed) such that even if the three considered obfuscators can produce focal-methods that are almost identical, a sample is considered unique if only one byte differs from any other sample.

\begin{figure}[t]
\centering
\input{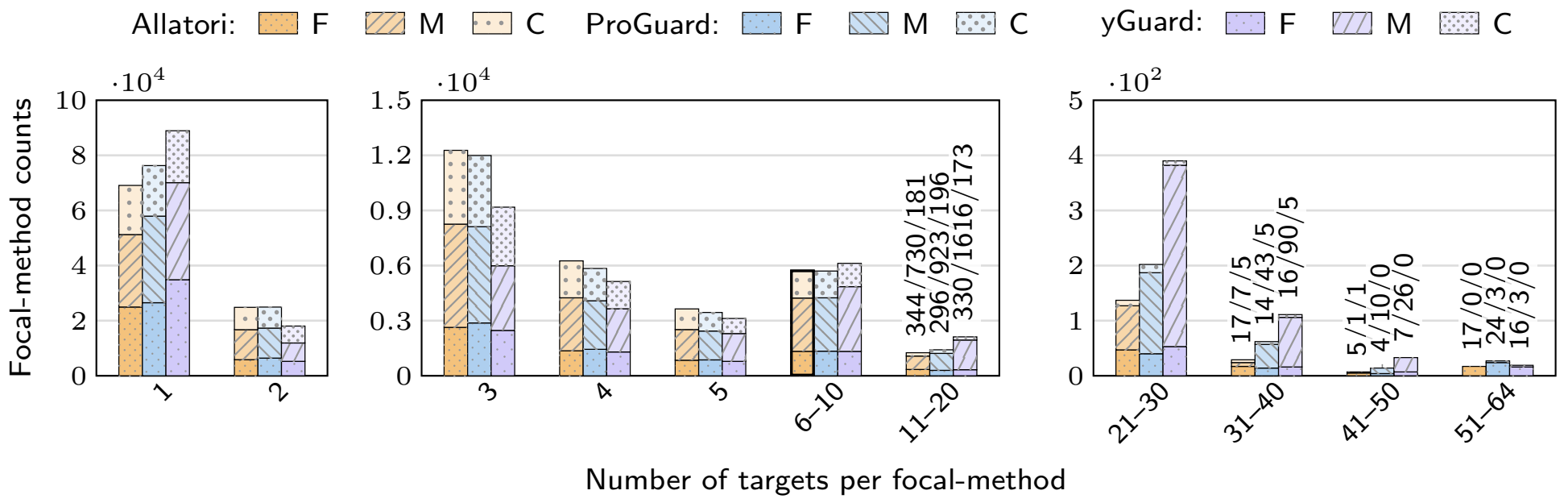}
\caption{Distribution of field (F), method (M), and class (C) targets per focal-method.
The panels use different linear y-scales.
Within each bin, bars show Allatori, ProGuard, and yGuard from left to right, with F, M, and C stacked from bottom to top.
Bins are computed separately for each identifier type: each segment counts focal-methods whose number of distinct targets of that type falls in the bin.
Small-stack annotations report these method counts in F/M/C order. Example: the outlined Allatori stack illustrates \num{1331} methods with 6--10 field targets, \num{2900} with 6--10 method targets, and \num{1520} with 6--10 class targets.}
\label{fig:target-count-histogram}
\end{figure}

\autoref{tab:identifier_target_stats} summarizes the final dataset statistics.
{For each obfuscator, it reports the number of unique focal-methods and the number of distinct targets in $T(m)$, counted once per focal-method and broken down by identifier type.}
These values remain consistent across the different context settings and evaluated models.
When several focal-methods use one target, their inventories contain the same private mapping ID for it. After deduplicating mapping IDs across focal-methods within each obfuscator, we obtain 307,097 obfuscator--target pairs in total across the three obfuscators.
\autoref{fig:target-count-histogram} further shows the target-count breakdown of our focal-methods for each obfuscator.
It shows that many more focal-methods comprise few targets.
The complete list of repositories, their creation dates, and the commits used for evaluation is provided in the supplementary material at \url{https://doi.org/10.6084/m9.figshare.34572789.v1}.

\textbf{Identifier obfuscation.}
We applied each obfuscator to the same compiled project JARs under a renaming-only configuration rather than its unrestricted defaults.  
\begin{itemize}
  \item \textbf{ProGuard~7.7.0} ran with \texttt{-dontshrink}, \texttt{-dontoptimize}, and \texttt{-dontpreverify}. It preserved structural attributes and used \texttt{-adaptclassstrings} to update class names embedded in string literals.
  \item \textbf{yGuard~5.0.0} executed only its \texttt{rename} task. It used the \texttt{small} naming scheme and generated Java-compatible names. We disabled name scrambling and overload-based name reuse but enabled replacement of class names in string literals.
  \item \textbf{Allatori~9.8 Educational} used \texttt{compact} names for classes, methods, and fields, and \texttt{abc} names for packages.
\end{itemize}
Notably, the obfuscators' default renaming strategy is to generate short, meaningless names, often one or two characters long, which are not semantically informative.
In this study, we chose to keep this default behavior as it is likely the most common scenario for a user of ad hoc tools.
Granted, a more complete experiment campaign would mix different identifier renaming strategies (\eg, high-entropy long names, or dictionary-based names) to assess the robustness of the evaluated approaches, but this is left for future work.
In \autoref{sec:results}, we nonetheless record the LLMs' inferred identifier-obfuscation strategy for each sample as a byproduct of the detection-task to give a broad picture of the LLMs' sensitivity to discrepancies in naming conventions.
Indeed, the LLMs are not given any information about the obfuscation strategy used by the obfuscator, and they don't actually know the snippet is unequivocally obfuscated (this fact is not included in the system or user prompt), but they are asked to identify this strategy.

For the main experiments we turned off string encryption, control-flow obfuscation, resource rewriting, member reordering, and other structural transformations. We examine compound obfuscation in \autoref{sec:ablation}.
We configured the obfuscators to preserve structural metadata, including generic signatures (\texttt{Signature}) and class-nesting information (\texttt{InnerClasses} and \texttt{EnclosingMethod}), for decompilation and analysis.
Allatori was also configured to keep local-variable names and line numbers; ProGuard and yGuard did not explicitly preserve these debug tables.

For each repository and obfuscator, and by tapping the obfuscation process, we construct a deterministic mapping between project-owned class, method, and field identifiers and the synthetic, meaningless names generated during that process.  
We saved the clean-to-obfuscated mapping separately as private ground truth, its only use in the study.

\subsection{Overall Workflow}

\begin{figure}[t]
  \centering
  \makebox[\linewidth][c]{%
    \includegraphics[width=\linewidth]{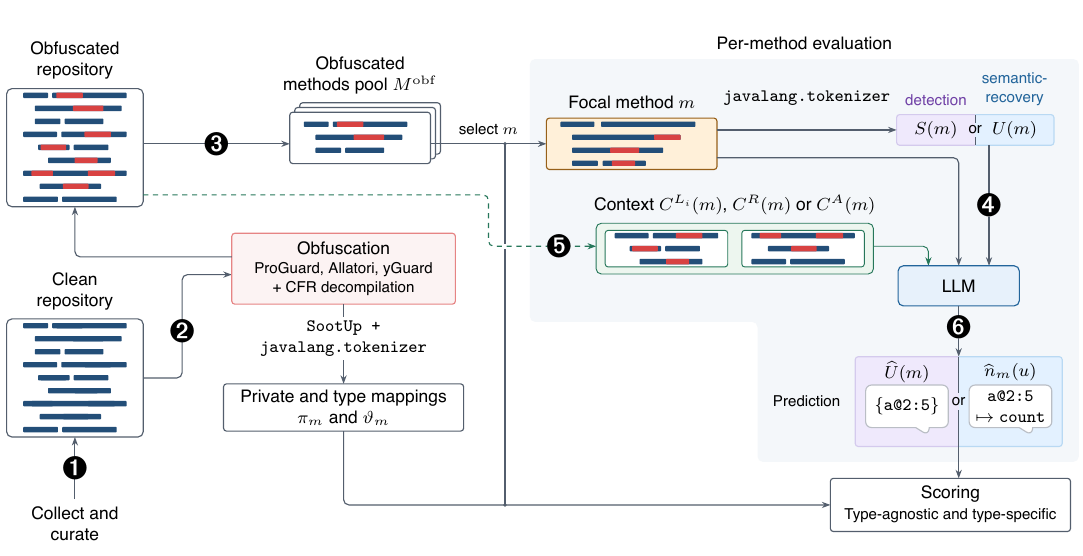}%
  }
  \caption{Overall workflow of the LLM evaluation. \ding{205} depends on the task, \ding{206} varies with the setting.}
  \label{fig:overall-workflow}
\end{figure}

\autoref{fig:overall-workflow} describes all the steps for evaluating the LLMs for either task.
We first collect and curate a set of open-source Java repositories selected to reduce training-data overlap (this step, \autoref{fig:overall-workflow}--\ding{202}, is described in detail in \autoref{ssec:data-collection}).
We then obfuscate the curated repositories with ProGuard, Allatori, and yGuard, and decompile the obfuscated bytecode with CFR to obtain the obfuscated source code (\autoref{fig:overall-workflow}--\ding{203}); this step also produces the mappings: $\pi_m$ to link occurrences to targets' clear and obfuscated names, and $\vartheta_m$ to get the targets' types.
We then extract the focal-methods, which are the source code of the methods we want to deobfuscate, from the decompiled obfuscated source code: they constitute the focal-methods pool (\autoref{fig:overall-workflow}--\ding{204}). 
The first three steps are the setup of our main experiment campaign so they are performed only once.

{During an experiment campaign, each focal-method $m$ is enriched with an occurrence inventory $S(m)$ or $U(m)$, obtained by the \texttt{javalang} lexer~\cite{javalang} and accompanied by row and column positions within the focal-method (\autoref{fig:overall-workflow}--\ding{205}).}
{For the detection task, $S(m)$ deliberately includes variables, parameters, packages, readable APIs, and other negatives so inventory membership does not reveal which occurrences belong to $U(m)$.}
{The detector must identify the occurrences of obfuscated project-owned fields, methods, and classes, producing $\widehat{U}(m)\subseteq S(m)$.}

{The focal-method $m$ and its task-specific inventory constitute a sample.}
{These are two of the three elements that enter the user prompt, the third being a piece of context which is drawn from the decompiled obfuscated source code of the project the sample belongs to: clear code never enters the evaluated approaches, be they traditional identifier-detection techniques or LLMs.}
This context extraction process (\autoref{fig:overall-workflow}--\ding{206}) is what separates our different settings: $L_0$ means context is null, $L_i$, $i=1,2,3$ have the chunks procedurally picked in various ways/richness, $R$ uses a retrieval mechanism and $A$ uses an agentic harness.

{For the detection task, the LLM classifies every occurrence in $S(m)$ as obfuscated or not, producing $\widehat{U}(m)$ (\autoref{fig:overall-workflow}--\ding{207}).}
The LLM also infers the type of each flagged occurrence (field, method, class): this is not used for scoring (not shown in the figure).
We request these types in every setting only because we found that it stabilized output accuracy (identifier types are not scored).
A reverse engineer is likely not interested in the intermediate goal of type-aware identifier detection, so this type-inference does not count towards any scoring: yet, we found that requesting the LLM to infer the type of each flagged identifier was beneficial to the model's ability to detect obfuscated identifiers.
{We finally resolve the flagged occurrences to target identities using the $\pi_m$ map and apply the \quotes{touch-any} rule to score detection against $T(m)$.}

{For semantic recovery, the LLM receives the focal-method $m$, the associated context, and the ground-truth target-occurrence inventory $U(m)$ (\autoref{fig:detection-semantic-recovery}).}
{Each entry identifies an occurrence by its spelling and syntax coordinates and supplies its identifier type; private mapping IDs and original names are withheld from the model.}
{The model predicts one name $\widehat{n}_m(u)$ for each $u\in U(m)$.}
All in all, steps \ding{205}, \ding{206}, and \ding{207} are how each method from $M^{\text{obf}}$ is evaluated for each task, each setting.

{In this paper, the two tasks are evaluated and scored separately: end-to-end identifier-recovery quality is evaluated only in the ablation exploration of \autoref{sec:ablation}, where the recovery inventory is built from the detector output $\widehat{U}(m)$ and inferred types, including possible false positives, instead of the ground-truth occurrence inventory $U(m)$.}
{This ablation measures the effect of missed true targets $T(m)\setminus\widehat{T}(m)$, where $\widehat{T}(m)=\pi_m(\widehat{U}(m)\cap U(m))$.}

\subsection{Retrieval Workflow}
\label{ssec:retrieval-workflow}

\begin{figure}[t]
  \centering
  \makebox[\linewidth][c]{%
    \includegraphics[width=\linewidth]{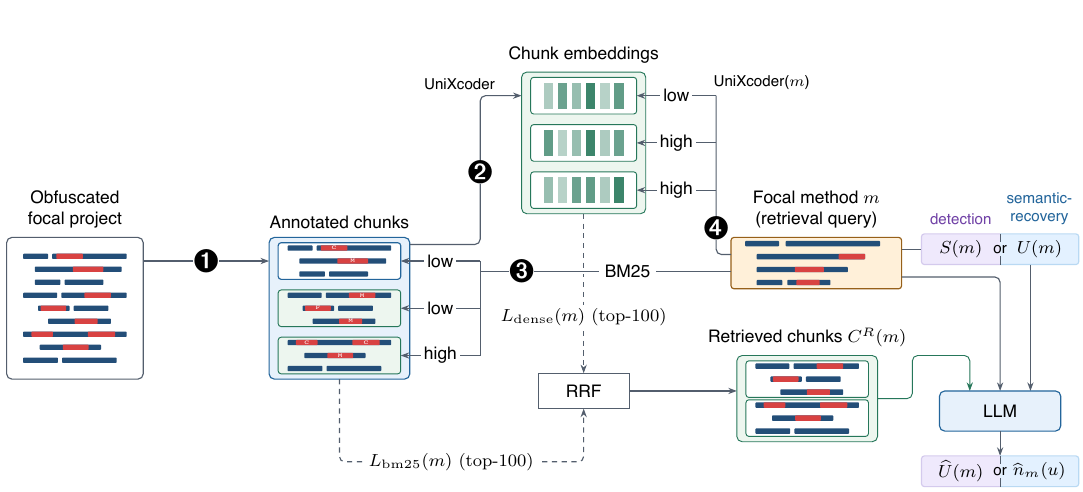}%
  }
  \caption{Retrieval workflow for the $R$ experiments. The focal project is the repository $m$ originates from.}
  \label{fig:retrieval-workflow}
\end{figure}

\autoref{fig:retrieval-workflow} describes the workflow of the $R$ experiments.
Retrieval operates exclusively on the CFR-decompiled obfuscated repository produced by the same obfuscation run.  
The retriever can never access clean source code or obfuscation mappings.
During preprocessing, we split the decompiled repository into code chunks.
The chunking rules encompass method bodies, caller/callee snippets and classes \etc
Each chunk is indexed by both BM25 and FAISS using the same identifier to allow their rankings to be matched and combined through reciprocal-rank fusion (see \autoref{ssec:settings}).
Tree-sitter identifies identifier roles, which are encoded as BM25 terms such as \texttt{identifier:method:b}; the stored source code remains unchanged (\autoref{fig:retrieval-workflow}--\ding{202}).
These role terms are used for retrieval and are not included in the code shown to the LLM.

In parallel, UniXcoder encodes each chunk together with its parser annotations into a vector stored in a FAISS index (\autoref{fig:retrieval-workflow}--\ding{203}).
At retrieval time, the obfuscated focal-method $m$ queries BM25 and UniXcoder/FAISS (\autoref{fig:retrieval-workflow}--\ding{204} and \ding{205}, respectively), each returning a ranked list of up to 100 repository chunks. 
{The task inventory (\ie, \(S(m)\) for detection or \(U(m)\) for semantic recovery) is not used by the retriever and is passed only to the downstream LLM.}

The BM25 and UniXcoder rankings are combined through reciprocal-rank fusion (\autoref{fig:retrieval-workflow}).  
This combines the two rankings without comparing scores measured on different scales.
Duplicate results are discarded and retrieved chunks are included in RRF order; chunks exceeding the remaining context budget are skipped rather than truncated.

Finally, the LLM receives the focal-method, the task-specific inventory, and the selected repository evidence, $C^R(m)$.  
Names appearing only in retrieved context can help a prediction. They do not become targets for that instance.

Our trials showed confusion when the detector was given role labels, so we do not include them in \(S(m)\): they only exist in the annotated chunks used for the retrieval, as shown in \autoref{fig:retrieval-workflow}--\ding{202}.
Each entry in \(S(m)\) is solely a candidate occurrence identified by its position in the focal-method.  
On the other hand, asking the model to return the inferred type of each occurrence that it flags improved its ability to detect obfuscated identifiers.
This is why we underline that identifier types are requested as part of the model's output but are not supplied in \(S(m)\) as input, nor are they considered for scoring.

\subsection{Agentic Workflow}
\begin{figure}[t]
  \centering
  \includegraphics[width=\linewidth]{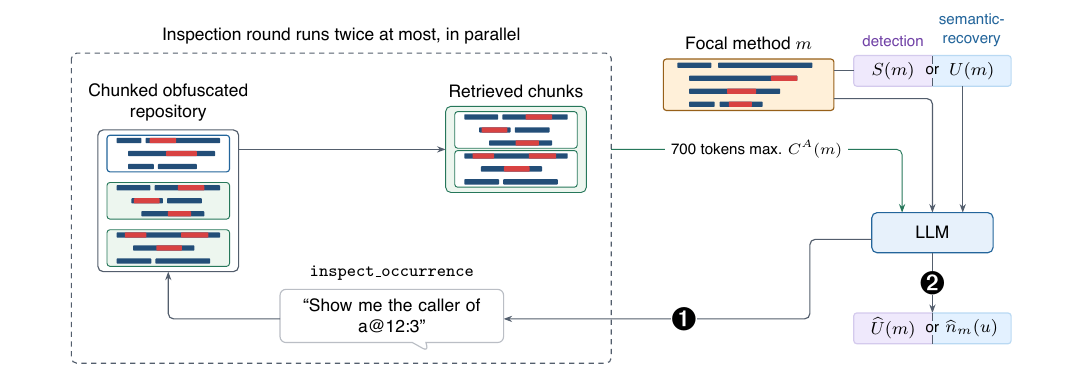}
  \caption{Agentic workflow for the $A$ experiments.}
  \label{fig:agentic-workflow}
\end{figure}

Figure~\ref{fig:agentic-workflow} describes the workflow of the $A$ experiments.
Each run on a focal-method \(m\) comprises two model requests: retrieval planning and final prediction.
{Given \(m\) and the inventory (all occurrences \(S(m)\) for detection, or typed occurrences \(U(m)\) for semantic recovery), the model requests one or two inspections of identifier occurrences (\autoref{fig:agentic-workflow}--\ding{202}).}
Our harness only has one available tool, which is read-only: \texttt{inspect\_occurrence}.
Each call specifies one inventory occurrence through \texttt{occurrence\_index} and one \texttt{facet}: \texttt{definition}, \texttt{owner}, \texttt{callers}, \texttt{callees}, or \texttt{field\_uses}.
For example, the following logged call asks for information about the owner of the occurrence indexed by 1:
\begin{quote}
  \small\ttfamily
  \{"tool": "inspect\_occurrence",
  "occurrence\_index": 1, "facet": "owner"\}
\end{quote}
and the harness maps $m$ to the obfuscated repository it originated from to retrieve the owner class.
Both calls are selected before any evidence is returned; therefore, the second call does not depend on the result of the first.
A single call retrieves obfuscated context up to 700 tokens, whereas two calls return 350 tokens each.
Each retrieved obfuscated code chunk given by the harness contains as many evidence blocks as possible and discards the overflowing block.
The combined retrieved evidence $C^A(m)$ is therefore limited to 700 tokens.

Whether one or two inspections are requested is up to the model itself.
It makes this choice based on the focal-method and the identifier inventory before seeing any context retrieved by the tool call.
It would request one inspection when a single occurrence appears to be the most useful ambiguity to study and two when two distinct symbol groups appear worth exploring.

The harness executes these inspections over the obfuscated repository in one round.
It excludes the focal-method from the retrieval.
The resulting evidence constitutes $C^A(m)$ and is appended to the user prompt; retrieval is then closed, and the model produces its final response (\autoref{fig:agentic-workflow}--\ding{203}) using $m+C^A(m)$ and either $S(m)$ or {$U(m)$}.
Before scoring, we save the tool requests, the evidence returned and the token usage.

\subsection{Traditional baselines}
\label{ssec:traditional-baselines}
We describe how we used or reimplemented the traditional baselines marked as used in Table~\ref{tab:deobfuscator_scope}.

\textbf{Isobe and Tamada.} This technique represents each focal-method by occurrence counts of grouped JVM opcodes and trains a random forest to predict the verb component of its original method name. 
The published system replaces only the leading verb of method names (\eg, \texttt{validate} in \texttt{validateInput}) using a fixed vocabulary; it does not recover complete method names or rename fields, classes, or other identifier types.
We could not obtain the authors' exact implementation, trained model, or vocabulary. We divided our 51 repositories into five disjoint folds. To evaluate a fold, we train on histograms from the clean JARs of the other four folds and test on histograms from its obfuscated JARs, removing training methods duplicated in the test fold beforehand. 
For this experiment, we do not care about temporality because we assume the time span is not long enough to affect the distribution of method-name verbs in the training and test folds.
Following the paper, training excludes constructors, methods outside the 30--1,000-instruction range, and the verbs \texttt{get} and \texttt{set}; the 20 most frequent remaining leading verbs are selected independently within each training fold.

\textbf{Anti-ProGuard.} Anti-ProGuard expects APK inputs exclusively. It searches a database of unobfuscated applications for bytecode resembling the obfuscated application using instruction \(n\)-grams and 256-bit SimHash fingerprints, and transfers names from accepted matches~\cite{DBLP:conf/dais/BaumannPM17}.
Its released renaming pipeline covers packages, classes, and methods, but not fields.
We use the released implementation at commit \texttt{f609cb1c}.
The released artifact had no populated fingerprint database. We built one with its original extractor and the \num{1684} clean F-Droid training APKs.
We convert each benchmark JAR to DEX without optimization, place it in a minimal APK, and submit the complete application to Anti-ProGuard.
Preparing this retrieval system means building the database; there are no parameters to train.
We ran this deobfuscator on 50 APKs obtained by passing our 51 obfuscated JAR files to D8 9.4.14 (one failed) with parameters 
\texttt{--release --no-desugaring --min-api 35}.

  \textbf{DeGuard.}
 DeGuard builds a graph that connects packages, classes, methods, and fields across the application, then predicts their names together while enforcing Java naming constraints~\cite{DBLP:conf/ccs/BichselRTV16}. 
  We could not obtain the original implementation, model, or exact training corpus. Our reimplementation uses a Soot-Jimple frontend and follows the original paper's logic.
  For training, we use the \num{1684} clean F-Droid applications that also populate the Anti-ProGuard database.
  Training traverses the applications once to learn which names tend to occur together in related code. At inference, the names of connected program elements are predicted together to keep them compatible.
  The complete obfuscated application is processed before predictions are projected onto class, method, and field occurrences in the focal-methods.
  This reimplementation of DeGuard was run on the same cohort of 50 APKs as Anti-ProGuard.

\textbf{DOBF.}  
DOBF uses an encoder--decoder Transformer to reconstruct class, function, and variable names replaced with typed placeholders in Java or Python source code.  
For our study we used the authors' \texttt{dobf.pth} checkpoint released among their artifacts, made of a 12-layer encoder, a 6-layer decoder, 768-dimensional hidden states, and 12 attention heads, together with its RoBERTa byte-pair encoding (BPE) tokenizer and CodeGen preprocessing pipeline~\cite{DBLP:conf/nips/LachauxRSL21}.  
The model is used without further training.  
Candidate selection follows DOBF's AST-based preprocessing: it uses the Java AST to identify class, method and variable names (including fields and locals) in the complete CFR-decompiled file and replaces them with typed placeholders (\eg, \texttt{CLASS\_0}, \texttt{FUNC\_0}, \texttt{VAR\_0}).
DOBF also replaces readable names of local variables and parameters with placeholders, so the model has fewer clues about what the code does.
The Transformer returns its predicted names in a dictionary indexed by the placeholders.
Placeholders are linked back to declarations in the obfuscated source, where AST information distinguishes fields from local variables and parameters.
The same output serves both evaluations: any nonempty decoded name counts as a rename attempt for detection under our \quotes{touch-any} rule (see \autoref{ssec:models-and-metrics}), even if it repeats the obfuscated spelling. Missing or empty entries do not count as attempts. Semantic recovery separately evaluates the proposed name against the original name.
We ran DOBF on all \texttt{.java} files of our 51 obfuscated repositories.
Only after inference do we use the private map $\pi_m$ to associate genuine target occurrences with their targets in $T(m)$ for scoring. DOBF's preprocessor does not use this private mapping.

\section{Results}
\label{sec:results}

\subsection{Obfuscation strategy}

{
Besides identifying individual targets (but still during the detection task), we ask the LLM to indicate whether the focal-method appears obfuscated and which naming strategies it recognizes: short-generated names, high-entropy names, or other opaque names.
We found that this request helped the inference.
These judgments give a small assessment of obfuscation at method level. Recognizing the strategy does not tell us whether the model found every affected identifier.

Effectively, there are no focal-methods inside our dataset that contain zero targets.
We consequently don't intend for the model to output \textit{None} because it implies the LLM considers $m$ as unobfuscated (no identifier within $S(m)$ is flagged).
The only scheme used for identifier renaming is \textit{Short-generated} as we use the default parameters of the obfuscators in this regard.
Consequently, we would expect a good model to output \textit{Short-generated} all the time.
Nonetheless, we may also consider \textit{Mix} a somewhat acceptable answer: it indicates that the model recognizes several strategies coexisting and should therefore approach the method with caution instead of bluntly assuming the obfuscation employed a single naming scheme.
}

\begin{figure*}[t]
  \centering
  \includegraphics[width=\textwidth]{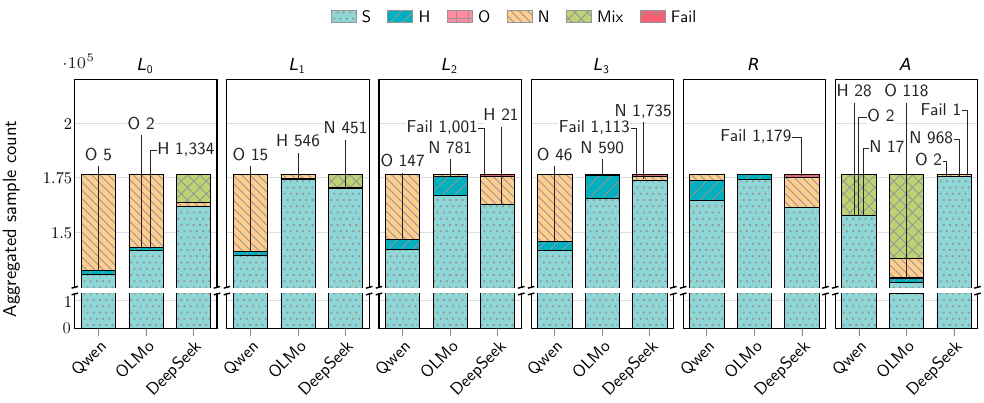}
  \caption{Inferred obfuscation families by model and context setting. Each stacked bar sums Allatori, ProGuard, and yGuard samples; segments show short-generated (S), high-entropy (H), other opaque (O), no inferred family (N), mixed naming strategies (Mix), and failed responses (Fail).}
  \label{fig:obfuscation-family-counts}
\end{figure*}

Figure~\ref{fig:obfuscation-family-counts} displays the inferred obfuscation families across Allatori, ProGuard, and yGuard for each model and context setting. 
With only the focal-method ($L_0$), Qwen and OLMo would often fail at recognizing the obfuscation family given they return \textit{none} for 28.9\% and 17.9\% of methods, respectively.
Additional context reduces this share to 15.8\% for Qwen under $L_2$ and down to 1.0\% for OLMo under only $L_1$.
More context does not always improve the prediction of the family, though: OLMo increasingly selects the high-entropy category under $L_2$ and $L_3$, assigning 10,303 methods to this category under $L_3$.
Fixed retrieval setting ($R$) nearly eliminates any \textit{none} answers from Qwen (1.5\%) and eliminates it completely for OLMo, whereas DeepSeek still assigns it to over 13,000 methods.
Agentic retrieval ($A$) produces more mixed answers for Qwen and OLMo (10.7\% and 21.8\%), which we acknowledge as recognition of obfuscation even though they do not identify the single applied scheme precisely.
With agentic retrieval, DeepSeek chooses short-generated in 99.4\% of responses; it returns \textit{none} 968 times, and one response fails.
The models generally recognize obfuscation better with additional context, although retrieval does not affect them alike. Under $A$, Qwen and OLMo more often infer mixed strategies; DeepSeek almost always recognizes the expected short-generated scheme.

\autoref{fig:obfuscation-family-counts} shows DeepSeek sometimes returns failed responses in detection. 
For Qwen and OLMo, generation is constrained by JSON schemas that require a complete inventory (\ie, the cardinality of $\widehat{U}(m)$ must be respected) that DeepSeek's API doesn't enforce (checks are conducted after receiving the API response).
These failures are equivalent to malformations for the detection task; we did not retry or repair them: the corresponding samples are discarded for DeepSeek on the detection task and, as such, do not count towards scoring this model in \autoref{ssec:detection-results} and \autoref{ssec:target-counts-vs-detection}.

\subsection{Detection under different settings}
\label{ssec:detection-results}
Table~\ref{tab:detection_scores_method_macro} reports the identifier-detection $F_1$ scores for the different experimental settings and models as well as type-specific scores.
A high value in this table means the LLM (or traditional tool) is good at correctly flagging an identifier of the corresponding type (field, method, class or altogether) as being obfuscated.

\textit{{Models and context.}}
Detection performance varies largely across models, context settings, identifier types, and obfuscators.
DeepSeek gives the highest overall detection scores. Even when its input is only the focal-method (\(L_0\)), total \(F_1\) is \(80.69\)--\(85.75\%\). At \(L_0\), classes appear harder for DeepSeek: we get \(65.27\%\) to \(72.78\%\) for \(F_{1,C}\), compared with \(77.87\%\) to \(83.54\%\) for fields (\(F_{1,F}\)) and \(83.52\%\) to \(89.05\%\) for methods (\(F_{1,M}\)). Surprisingly, adding static context (\(L_1\)--\(L_3\)) reduced DeepSeek's overall detection \(F_1\) for all three obfuscators. With the full enclosing class (\(L_2\) setting), scores were 16.25--20.29 percentage points lower than with the focal-method alone (\(L_0\)).
We hypothesize that additional context can introduce irrelevant code that interferes with identifier detection. 
Similar effects have been reported for repository-level code completion, where retrieved context can degrade performance \cite{WuEtAl2024Repoformer}, and for longer inputs even when relevant information is brought in \cite{DuEtAl2025ContextLength}. 
These separate works provide possible explanations for DeepSeek's lowering results but we do not formally identify the cause.

OLMo improved mainly on methods and fields from \(L_0\) to \(L_1\), gaining up to 18.37 points in \(F_{1,M}\) and 13.01 in \(F_{1,F}\). For classes, the gain in \(F_{1,C}\) never exceeded 2.53 points. With this moderate context (\(L_1\)), we obtained its best static total \(F_1\): \(60.52\)--\(73.18\%\). Qwen was considerably weaker across \(L_0\)--\(L_3\), but reached \(31.01\)--\(38.42\%\) total \(F_1\) with RAG retrieval (\(R\)). Compared with \(L_0\), \(R\) also helped OLMo, but substantially reduced DeepSeek's performance.

\textit{{Agentic retrieval.}}
The largest gains are with \(A\). Qwen reaches \(80.85\)--\(87.40\%\) total \(F_1\); DeepSeek reaches \(87.29\)--\(91.56\%\), and OLMo \(72.77\)--\(80.37\%\). We find much higher field and method scores than class scores under \(A\): \(F_{1,F}\) is between \(93.31\%\) and \(98.58\%\), and \(F_{1,M}\) between \(86.99\%\) and \(98.25\%\), compared with \(43.83\%\) to \(76.68\%\) for \(F_{1,C}\). DeepSeek is the best at detecting classes in \(A\), scoring \(67.43\%\) to \(76.68\%\) in \(F_{1,C}\). Classes nevertheless remain its main bottleneck.

\textit{{Traditional baselines.}}
Anti-ProGuard and the DeGuard baseline score near zero across the types they support. Isobe--Tamada remains competitive for methods (\(F_{1,M}\) of \(69.79\%\) to \(75.01\%\)), but covers neither fields nor classes, as discussed in \autoref{ssec:traditional-baselines}.

DOBF is the best general baseline, at \(43.83\%\) to \(45.83\%\) total \(F_1\). Its best type is classes (\(F_{1,C}\) of \(61.19\)--\(69.29\%\)), unlike the LLMs, yet its aggregate results are far below DeepSeek's. DeepSeek exceeds it by roughly 36--42 percentage points of total \(F_1\) with \(L_0\), and about 43--48 points with \(A\).

Across all LLM models, settings, and obfuscators, class identifiers resist detection most: mean \(F_{1,C}\) is \(48.48\%\). This compares with \(61.22\%\) for fields (\(F_{1,F}\)) and \(62.78\%\) for methods (\(F_{1,M}\)); class detection is lower by \(12.74\) and \(14.30\) percentage points on average, respectively.

It is important to adjust the appreciation of the three models' performances with respect to their possible prior exposure to the focal-methods of our obfuscated repositories: while we construct our clear dataset to eliminate strict contamination for OLMo thanks to the provided pretraining corpus, we cannot strictly guarantee this for Qwen or DeepSeek.
Even in that regard, OLMo remains a better fit for the detection task than the traditional baselines performance wise.

\begin{sidewaystable}[!p]
    \centering
    \caption{\textit{Higher is better}. Type-aware (columns F, M and C) and type-agnostic (\quotes{Total}) $F_1$-scores (\%) of obfuscated-identifier detectors.
    Scores are reported as method-macro averages (58,065 methods for Allatori, 58,721 for ProGuard, and 59,078 for yGuard).
    The bottom block reports four traditional baselines on their native settings. \quotes{-} symbols mean not in scope for the deobfuscator. Bold values indicate the best score within the LLM block.}
    \label{tab:detection_scores_method_macro}
    \scriptsize
    \setlength{\tabcolsep}{2.4pt}
    \renewcommand{\arraystretch}{1.05}
    
\newcommand{\semanticDeobfuscationScoreRows}{%
\multicolumn{2}{l}{Isobe--Tamada}
& -- & -- & --
& 69.79 & 73.58 & 75.01
& -- & -- & --
& 38.63 & 42.96 & 40.29 \\
\multicolumn{2}{l}{Anti-ProGuard}
& -- & -- & --
& 0.00 & 0.00 & 0.00
& 0.00 & 0.00 & 0.00
& 0.00 & 0.00 & 0.00 \\
\multicolumn{2}{l}{DeGuard}
& 0.37 & 0.95 & 0.55
& 1.03 & 1.20 & 1.63
& 0.00 & 0.00 & 0.00
& 0.50 & 1.01 & 0.88 \\
\multicolumn{2}{l}{DOBF}
& 57.49 & 48.14 & 50.03
& 55.01 & 48.65 & 47.46
& 69.29 & 61.62 & 61.19
& 45.83 & 44.26 & 43.83 \\
}

    \begin{tabular*}{\textheight}{@{\extracolsep{\fill}}ll*{12}{c}@{}}
        \toprule
        \textbf{Setting} & \textbf{Model}
        & \multicolumn{3}{c}{\textbf{F}}
        & \multicolumn{3}{c}{\textbf{M}}
        & \multicolumn{3}{c}{\textbf{C}}
        & \multicolumn{3}{c}{\textbf{Total}} \\
        \cmidrule(lr){3-5}
        \cmidrule(lr){6-8}
        \cmidrule(lr){9-11}
        \cmidrule(lr){12-14}
        & & Allatori & ProGuard & yGuard
        & Allatori & ProGuard & yGuard
        & Allatori & ProGuard & yGuard
        & Allatori & ProGuard & yGuard \\
        \midrule
        $L_0$ & Qwen & 31.47 & 16.82 & 20.99 & 26.84 & 16.25 & 22.68 & 25.39 & 15.72 & 16.78 & 26.63 & 16.04 & 20.13 \\
         & OLMo & 55.44 & 54.59 & 57.87 & 59.66 & 50.49 & 62.86 & 55.77 & 51.67 & 59.83 & 53.76 & 46.74 & 57.24 \\
         & DeepSeek & 80.45 & 77.87 & 83.54 & 83.84 & 83.52 & 89.05 & 72.78 & 65.27 & 71.13 & 83.02 & 80.69 & 85.75 \\
        \midrule
        $L_1$ & Qwen & 33.12 & 21.07 & 26.36 & 31.76 & 24.85 & 29.67 & 28.03 & 18.83 & 20.01 & 30.73 & 22.95 & 26.84 \\
         & OLMo & 59.51 & 65.53 & 70.88 & 68.18 & 64.41 & 81.23 & 56.43 & 54.20 & 61.79 & 62.27 & 60.52 & 73.18 \\
         & DeepSeek & 77.74 & 79.02 & 85.28 & 83.06 & 79.99 & 87.14 & 71.39 & 61.47 & 67.81 & 81.51 & 77.78 & 84.40 \\
        \midrule
        $L_2$ & Qwen & 33.27 & 19.16 & 25.21 & 32.88 & 22.72 & 28.17 & 28.38 & 18.03 & 19.80 & 31.90 & 21.38 & 25.98 \\
         & OLMo & 54.06 & 58.49 & 69.46 & 68.17 & 65.86 & 78.75 & 55.45 & 51.60 & 56.86 & 60.86 & 59.64 & 71.11 \\
         & DeepSeek & 65.63 & 67.21 & 76.63 & 65.09 & 58.66 & 68.73 & 56.14 & 51.23 & 58.05 & 63.93 & 60.40 & 69.50 \\
        \midrule
        $L_3$ & Qwen & 30.83 & 16.66 & 23.63 & 31.42 & 21.44 & 28.12 & 26.72 & 16.42 & 18.33 & 30.26 & 20.01 & 25.12 \\
         & OLMo & 53.85 & 58.54 & 69.14 & 68.10 & 66.20 & 78.57 & 53.39 & 50.22 & 55.39 & 60.30 & 59.56 & 70.65 \\
         & DeepSeek & 68.08 & 67.12 & 78.81 & 71.44 & 63.05 & 74.23 & 59.70 & 53.78 & 62.06 & 70.67 & 64.76 & 75.42 \\
        \midrule
        $R$ & Qwen & 43.09 & 25.85 & 38.37 & 29.35 & 33.40 & 40.21 & 33.77 & 20.48 & 22.44 & 35.39 & 31.01 & 38.42 \\
         & OLMo & 56.95 & 60.46 & 68.64 & 71.44 & 68.22 & 79.92 & 52.49 & 47.47 & 54.55 & 63.19 & 61.12 & 71.86 \\
         & DeepSeek & 72.58 & 71.83 & 70.32 & 70.36 & 66.34 & 64.44 & 59.81 & 54.65 & 55.39 & 68.83 & 66.08 & 63.51 \\
        \midrule
        $A$ & Qwen & \textbf{97.30} & \textbf{96.96} & \textbf{98.58} & \textbf{90.68} & \textbf{92.52} & \textbf{98.25} & 65.68 & 51.98 & 60.05 & 85.21 & 80.85 & 87.40 \\
         & OLMo & 94.68 & 93.31 & 93.77 & 86.99 & 87.62 & 93.02 & 57.78 & 43.83 & 53.27 & 78.22 & 72.77 & 80.37 \\
         & DeepSeek & 96.96 & 95.81 & 97.35 & 90.19 & 92.47 & 97.79 & \textbf{76.68} & \textbf{67.43} & \textbf{74.34} & \textbf{88.73} & \textbf{87.29} & \textbf{91.56} \\
        \hline\hline
        \noalign{\vskip 3pt}
        \semanticDeobfuscationScoreRows
        \bottomrule
    \end{tabular*}
\end{sidewaystable}

\subsection{Influence of the target count on detection}
\label{ssec:target-counts-vs-detection}

\begin{figure}[t]
  \centering
  \includegraphics[width=\linewidth]{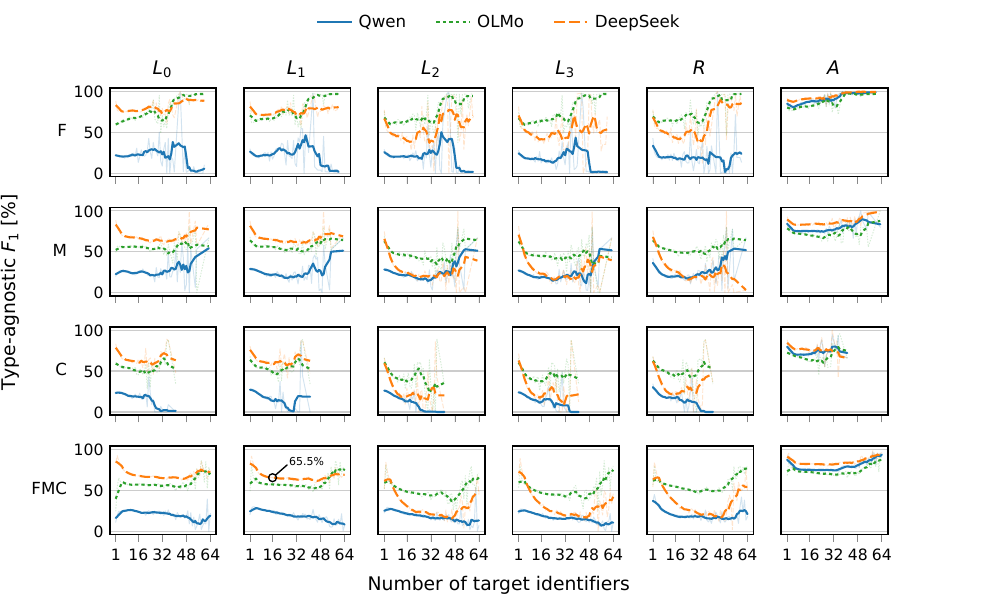}
  \caption{Mean type-agnostic detection $F_1$ by target count.
  {Scores are focal-method averaged; counts refer to distinct targets in $T(m)$.}
  Thick curves show seven-bin moving averages.
  Example: the circled point means DeepSeek scored 65.5\% under $L_1$ across 903 evaluated focal-methods containing exactly 16 targets (any type).}
  \label{fig:identifier-f1-by-target-count}
\end{figure}

{Table~\ref{tab:detection_scores_method_macro} displays both type-specific and type-agnostic detection performance with varying quantity of additional context. 
\autoref{fig:identifier-f1-by-target-count} shows only how type-agnostic detection $F_1$ varies with the number of distinct targets in $T(m)$ of a given identifier type in the focal-method, averaged over all obfuscators and focal-methods with the same target count.}
{Within each focal-method, target-level TP, FP, and FN counts are combined across all identifier types to compute $F_1$, then focal-methods are averaged with equal weight within each exact count bin.}
Rows specify which targets are counted on the horizontal axis: fields (F), methods (M), classes (C), or all targets (FMC); the vertical axis always reports aggregate detection $F_1$, and columns denote context settings.
Results are pooled across Allatori, ProGuard, and yGuard; all six settings use the same evaluation cohort of 175,864 focal-methods.

In our dataset, a higher number of targets can generally be construed as a longer focal-method, though it is not always the case (we found little difference experimentally in the downstream conclusions either way).
Notice that the curves are not of the same length between rows: this is because the maximum number of class targets per focal-method is not the same as the maximum number of fields per focal-method among all focal-methods of the dataset.

{
\textit{Target count and context.}
A higher target count does not reduce detection scores uniformly.
For DeepSeek, the clearest deterioration is in $L_2$, $L_3$, and $R$ as method, class, or total counts increase. In $L_3$, it scores $F_1$ of 52\% at eight total targets and 19\% at 32.
Aggregate detection also deteriorates with class count in several settings: for DeepSeek under $L_3$, $F_1$ falls from 71\% at one class target to 27\% at eight.

\textit{Model differences.}
OLMo is less sensitive than DeepSeek under these settings, but still declines: its $L_3$ score falls from 66\% to 39\% between one and eight class targets.
The field-count row shows a different pattern, with OLMo often improving towards the right, while Qwen remains weak and approaches zero in several high-count regions.
Even at $L_0$, OLMo rises from 59\% at one field target to 69\% at 16.
Several method-count and total-count curves also rise at the high-count end.

\textit{Agentic retrieval.}
$A$ maintains higher scores across most count bins and reduces the aforementioned declines seen for DeepSeek under $L_3$: its $F_1$ falls from 93\% to 79\% between one and 32 method targets, compared with 81\% to 17\% under $L_3$.
It remains sensitive to class count as it falls from 88\% at one class target to 77\% at eight.
The rebounds of the smoothed curve at high target counts call for caution, however: each $A$ field-count bin at 48 targets contains only three focal-methods, and the total-count bin at 64 only contains twelve.
}

\begin{figure}[t]
  \centering
  \includegraphics[width=\textwidth]{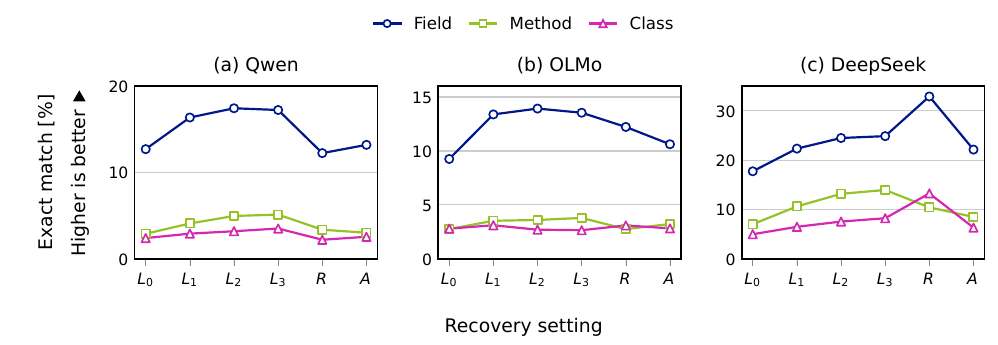}
  \caption{{Method-macro exact match recovery by identifier type across context settings, averaged with equal weight over evaluated focal-methods containing that type. Panels use different y-scales.}}
  \label{fig:semantic-recovery-by-kind-context}
\end{figure}

\subsection{Semantic recovery}
\label{ssec:semantic-recovery-results}
Independently from the detection task, we evaluate the ability of our considered LLMs and baselines to infer the semantics of a meaningless lexeme based on the surrounding obfuscated source-code context.
{\autoref{tab:semantic-recovery-pooled} shows type-agnostic similarity scores in our different considered settings, averaged with equal weight over evaluated focal-methods across the available obfuscators.}

{
\begin{sidewaystable}[!p]
    \centering
    \caption{{We pool identifier types to score semantic recovery.
    {For the LLMs, percentages are method-macro averages across the three obfuscators, with 175,864 focal-methods in the full cohort.}
    The traditional baselines use touched-method-macro averages, which leave out focal-methods without an emitted prediction. Bold values indicate the best score within the LLM block.}}
    \label{tab:semantic-recovery-pooled}
    \scriptsize
    \setlength{\tabcolsep}{2.5pt}
    \renewcommand{\arraystretch}{1.02}
    \begin{tabular*}{\textheight}{@{\extracolsep{\fill}}ll*{7}{r}@{}}
        \toprule
        \textbf{Setting} & \textbf{Model}
        & \textbf{ExM} & \textbf{LCS} & \textbf{LevSim} & \textbf{Jac}
        & \textbf{RL-$F_1$} & \textbf{B$F_1$} & \textbf{SemSim} \\
        \midrule
        $L_0$ & Qwen & 5.67 & 33.51 & 27.43 & 13.92 & 17.79 & 53.07 & 77.95 \\
         & OLMo & 4.63 & 30.77 & 25.46 & 12.68 & 16.31 & 49.72 & 78.89 \\
         & DeepSeek & 9.86 & 36.66 & 32.14 & 20.29 & 24.84 & 54.76 & 80.43 \\
        \midrule
        $L_1$ & Qwen & 7.49 & 35.45 & 29.40 & 16.34 & 20.38 & 54.51 & 77.75 \\
         & OLMo & 6.32 & 31.58 & 26.38 & 14.26 & 17.81 & 50.39 & 78.43 \\
         & DeepSeek & 13.50 & 40.45 & 36.07 & 24.87 & 29.72 & 57.65 & 80.93 \\
        \midrule
        $L_2$ & Qwen & 8.35 & 36.55 & 30.60 & 17.68 & 21.89 & 55.32 & 77.91 \\
         & OLMo & 6.47 & 31.77 & 26.43 & 14.16 & 17.62 & 50.93 & 77.45 \\
         & DeepSeek & 15.77 & 43.25 & 38.99 & 28.17 & 33.29 & 59.68 & 81.29 \\
        \midrule
        $L_3$ & Qwen & 8.42 & 36.99 & 31.02 & 18.11 & 22.45 & 55.64 & 77.95 \\
         & OLMo & 6.46 & 32.27 & 26.79 & 14.39 & 17.95 & 51.43 & 77.67 \\
         & DeepSeek & \textbf{16.42} & \textbf{44.39} & \textbf{40.13} & \textbf{29.33} & \textbf{34.61} & \textbf{60.44} & \textbf{81.49} \\
        \midrule
        $R$ & Qwen & 5.79 & 33.94 & 27.77 & 14.28 & 18.23 & 53.62 & 78.29 \\
         & OLMo & 5.48 & 32.48 & 26.67 & 13.81 & 17.53 & 51.54 & 76.01 \\
         & DeepSeek & 14.31 & 44.31 & 39.75 & 27.93 & 33.28 & 59.78 & 80.10 \\
        \midrule
        $A$ & Qwen & 5.90 & 33.08 & 27.06 & 13.63 & 17.24 & 52.81 & 78.74 \\
         & OLMo & 5.31 & 29.03 & 24.52 & 13.04 & 16.52 & 48.80 & 79.25 \\
         & DeepSeek & 12.09 & 38.33 & 34.09 & 22.48 & 26.83 & 55.68 & 80.25 \\
        \hline\hline
        \noalign{\vskip 3pt}
         & Isobe--Tamada & 5.25 & 22.23 & 13.52 & 5.25 & 5.25 & 48.69 & 99.73 \\
         & Anti-ProGuard & 0.00 & 16.65 & 11.85 & 0.68 & 1.11 & 44.60 & 67.58 \\
         & DeGuard & 1.14 & 11.49 & 9.70 & 3.30 & 4.35 & 39.31 & 82.97 \\
         & DOBF & 1.07 & 11.49 & 10.79 & 5.31 & 7.45 & 38.45 & 79.23 \\
        \bottomrule
    \end{tabular*}
    \par\smallskip
\end{sidewaystable}
}

\textit{{Static context.}}
Tables~\ref{tab:detection_scores_method_macro} and~\ref{tab:semantic-recovery-pooled} show that the most effective context setting depends on the task: $A$ achieves the highest total detection $F_1$ for every model and obfuscator, but $L_3$ leads most semantic-recovery metrics.
This distinction is noteworthy for DeepSeek, whose semantic-recovery scores improve steadily from $L_0$ to $L_3$, even though the same additional context reduced its scores for the detection task.
Qwen follows a similar pattern, with ExM increasing from $5.67\%$ at $L_0$ to $8.42\%$ at $L_3$, whereas OLMo gains little from increasing context beyond $L_1$.  

\textit{{RAG.}}
Repository retrieval ($R$) provides an acceptable alternative for DeepSeek: it ranks second on LCS, LevSim, and BERTScore $F_1$, just 0.08--0.66 percentage points below $L_3$.  
The benefits of $R$ depend on the model: OLMo obtains its highest LCS and BERTScore $F_1$ with retrieval, while Qwen performs better under $L_1$--$L_3$ on exact and lexical metrics.  

\textit{{Agentic retrieval.}}
By contrast, the strong results obtained under $A$ for the detection task do not extend to semantic recovery, where DeepSeek's ExM reaches only $12.09\%$, or 4.33 percentage points below $L_3$.  
The contrast suggests that detecting obfuscation and recovering meaning can benefit from different context.  
We relegate some detailed type-specific semantic-recovery scores to Appendix~\ref{app:semantic-recovery-by-kind}, Table~\ref{tab:semantic-recovery-by-kind}.
Figure~\ref{fig:semantic-recovery-by-kind-context} nonetheless provides a simple insight into the type-specific ExM score.
Broadly, it shows that fields remain easier to recover exactly than methods or classes. 
Qwen's $R$ and $A$ campaigns do not surpass the best static-context exact-match scores: $L_3$ for methods and classes, and $L_2$ for fields, where recovery peaks at 17.45\%.
OLMo $R$ trails $L_3$ for every identifier type, while $A$ trails $L_3$ for fields and methods and improves classes by only 0.17 points.
For DeepSeek, $R$ reaches 32.93\%, 10.43\%, and 13.26\% for fields, methods, and classes, while $A$ reaches 22.15\%, 8.49\%, and 6.31\%.

The greater resistance of class identifiers to recovery, compared with method and especially field identifiers, may stem from the more conceptual nature of class names, which often refer to abstractions defined in other modules and are therefore less directly inferable from the focal-method's logic, especially compared to field names whose logic is almost entirely and always contained from the get-go in the focal-method.

\textit{{Traditional baselines.}}
The earlier baselines also remain below DeepSeek on exact-match and lexical metrics.
In particular, Isobe--Tamada's $99.73\%$ SemSim is noteworthy but should be put in perspective in that this tool only predicts the leading action verb, which often carries most of a method name's meaning. For instance, \texttt{validate} for \texttt{validateConfiguration} can receive a high semantic similarity score even without the full identifier recovered.

\subsection{Influence of the target count on semantic recovery}

\begin{figure}[t]
  \centering
  \includegraphics[width=\linewidth]{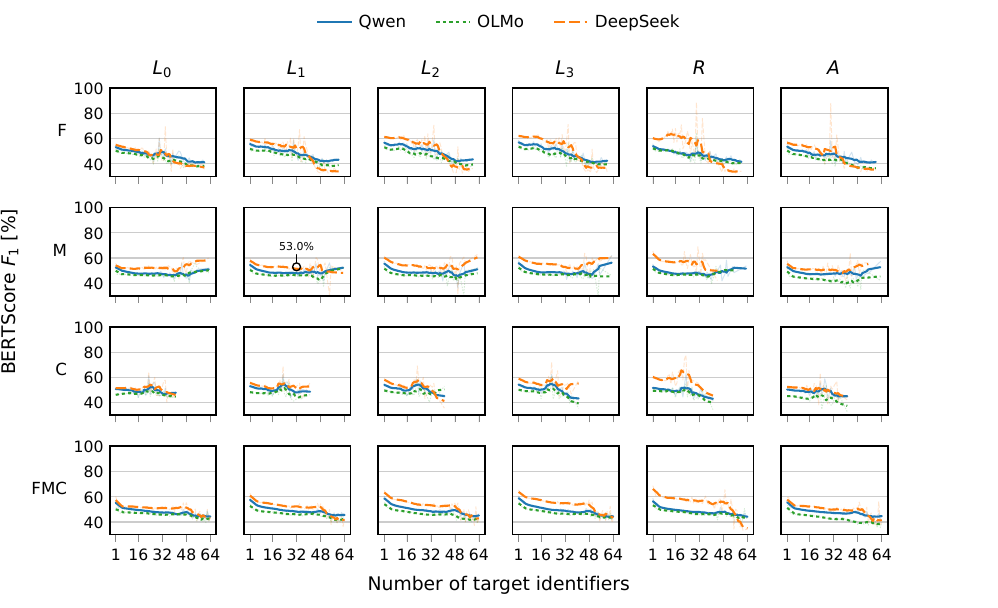}
  \caption{{Mean focal-method type-agnostic semantic-recovery BERTScore $F_1$ by distinct target count $|T(m)|$ or its type-specific subset.}
  Thick curves show seven-bin moving averages.
  Example: the circled point means DeepSeek scored 53.0\% under $L_1$ across 20 evaluated focal-methods containing exactly 32 method-identifier targets.}
  \label{fig:semantic-bf1-by-target-count}
\end{figure}

{{\autoref{fig:semantic-bf1-by-target-count} shows how semantic-recovery BERTScore $F_1$ varies with the number of distinct field, method, class, or total targets in $T(m)$.}}
Rows specify which targets are counted on the horizontal axis: fields (F), methods (M), classes (C), or all targets (FMC); the vertical axis always reports type-agnostic BERTScore $F_1$, and columns denote context settings.
Results are pooled across Allatori, ProGuard, and yGuard; all six settings use the same evaluation cohort of 175,864 focal-methods.

{
\textit{Target count and identifier type.}
We see a clear downward trend in semantic recovery as field and total target counts increase. Curves against method count are flatter after the initial decline, and sometimes rise again at high counts. With DeepSeek under $L_3$, eight field targets give BERTScore $F_1$ of 61\% and 32 give 49\%. For method targets, the change from eight to 32 is 57\% to 53\%. Grouping by class targets gives aggregate BERTScore $F_1$ of 55\% at eight and 52\% at 16. The class curves fluctuate, without a steady decline.
We should not read the rows as a ranking of recoverability by identifier type: they score all types together, and counts for the other types are allowed to vary.

\textit{Context and model differences.}
Moving from $L_0$ to $L_3$ at eight total targets gives DeepSeek the largest gain in aggregate BERTScore $F_1$, from 52\% to 59\%. The gains from richer static context are smaller for Qwen, from 51\% to 53\%, and OLMo, from 48\% to 50\%.
In the count range with more samples, DeepSeek generally keeps its lead with $L_2$ and $L_3$, differing from what we saw for detection.
For higher target counts, more context gives no consistent benefit; sometimes the models even exchange ranks.

\textit{Repository access.}
The strong agentic detection results in \autoref{fig:identifier-f1-by-target-count} do not really translate into a semantic-recovery advantage comparably.
At 32 total targets, OLMo gets BERTScore $F_1$ of 42\% with $A$ and 46\% with $L_3$.
The curves for DeepSeek with $R$ often come close to or above $L_3$.

Regions of the graphs with a high count of targets still call for caution: DeepSeek's $L_3$ field-count bin contains 685 evaluated methods at eight field targets but only 15 at 32; its total-count bin contains 5,133 methods at eight targets but only 12 at 64.
All in all, the curves support an association between target count and recovery difficulty, but don't necessarily imply that additional targets themselves cause the decline.
}

\bluebox{\textbf{\hypertarget{aq:1}{\hyperlink{rq:1}{AQ1}}:} LLMs largely outperform traditional baselines on the identifier recovery task, although it is not always the case: weaker models (in our experiments, Qwen) placed in an unfavorable setting ($L_0$) can perform worse than a simple Transformer (DOBF) or even a non-neural baseline (Isobe--Tamada) on detection.
LLMs are also more flexible regarding their attack surface, \eg, DOBF still requires an entire module for inference whereas the LLM can be fed smaller units.}

\bluebox{\textbf{\hypertarget{aq:2}{\hyperlink{rq:2}{AQ2}}:} Class identifiers are the hardest for LLMs to detect: even in the agentic setting, class detection remains well below field and method detection.
For semantic recovery, field names are easier to recover exactly for all three LLMs, while class names are the most resistant.
Richer additional context generally improves recovery without eliminating these differences.
The ranking depends on the metric and setting, \eg, DeepSeek under $R$ recovers more class names exactly than method names.
Semantic-recovery quality also tends to decline as focal-methods grow longer and richer in targets.}

\subsection{Prediction quality and inference cost}

To uncover insights from the LLMs' inference performances, we study their efficiency on the identifier recovery task.

\begin{figure*}[t]
  \centering
  \includegraphics[width=\textwidth]{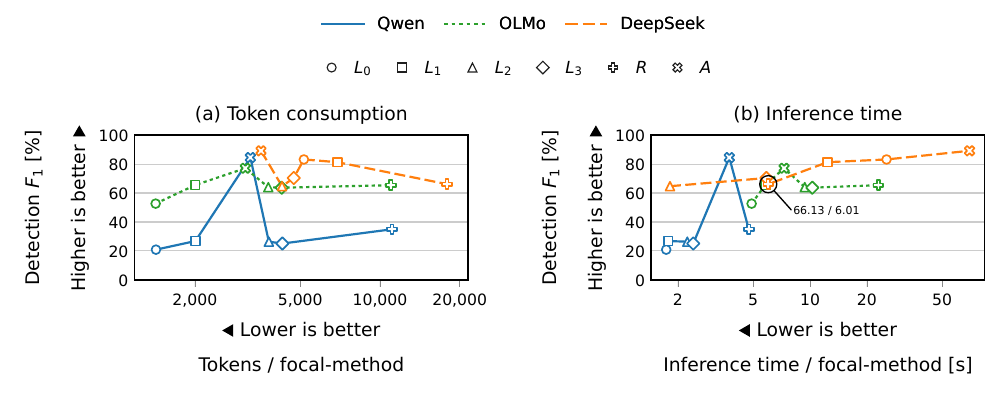}
\caption{Identifier-detection quality \textit{vs.} total token consumption (a) and mean inference time (b) per focal-method for our six settings.
  For scores and costs, we average over the same pool of focal-methods from Allatori, ProGuard, and yGuard.
  The connections between setting markers are arbitrary.
  Example: the circled $R$ point in (b) shows DeepSeek's mean detection $F_1$ of 66.13\% and mean inference time of 6.01\,s per focal-method, computed over the same 175,864 methods pooled across Allatori, ProGuard, and yGuard.}
  \label{fig:detection-resource-quality}
\end{figure*}

\begin{figure*}[t]
  \centering
  \includegraphics[width=\textwidth]{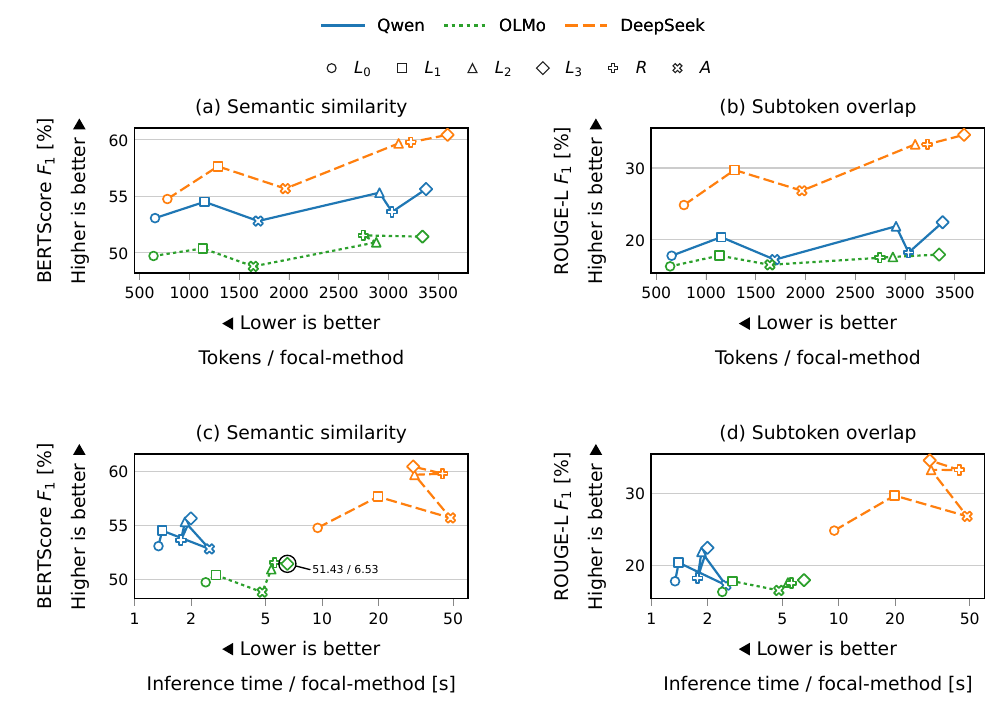}
  \caption{Semantic-recovery quality \textit{vs.} token consumption (a--b) and mean inference time (c--d) per focal-method for our six settings.
  Lines connect the setting markers arbitrarily.
  Example: the circled $L_3$ point in (c) shows OLMo's mean BERTScore $F_1$ of 51.43\% and mean inference time of 6.53\,s per focal-method. Quality values match Table~\ref{tab:semantic-recovery-pooled}; token counts and inference times summarize the saved run logs.}
  \label{fig:semantic-recovery-resource-quality}
\end{figure*}

\textit{Token consumption.}
Figure~\ref{fig:semantic-recovery-resource-quality}-(a--b) shows that even if the token burden of each static setting is similar across the three models, the quality of the output is not: DeepSeek gains 5.68 BERTScore $F_1$ and 9.77 RL-$F_1$ percentage points from $L_0$ to $L_3$; Qwen gains 2.57 and 4.66 percentage points, and OLMo gains 1.71 and 1.64 percentage points.
For detection, the additional tokens are less useful: moving from $L_1$ to $L_3$ roughly doubles token consumption for Qwen and OLMo without improving their $F_1$ (Figure~\ref{fig:detection-resource-quality}a).
As a reminder, detection receives the full candidate inventory \(S(m)\), whereas semantic recovery receives only \(U(m)\); this cardinality difference can incur significant token consumption between the two tasks.
The detection task additionally requires an inference on the type of the detected identifiers as a supporting sub-task (see \autoref{ssec:retrieval-workflow}).
Retrieved-context lengths also differ between tasks.
For both tasks, input token changes occur for three reasons:
\begin{itemize}
  \item each model uses its own tokenizer;
  \item because we cannot control DeepSeek’s generation through its API as precisely as we can for the locally hosted models (Qwen and OLMo), we add a short instruction in the user prompt describing the required response format.
  \item in the $A$ setting, the additional context $C^A(m)$ depends on the model's query (the retrieval of the $R$ setting is deterministic).
\end{itemize}
For semantic recovery specifically, output tokens may change only because the inferred names can be of varying length.
As a whole, larger static contexts increase token consumption for all three models, with the clearest recovery gains for DeepSeek and smaller improvements for Qwen and OLMo. 
Agentic retrieval reduces token consumption relative to $L_3$, but does not match its recovery quality.
For the detection task, however, $A$ both consumes 24--27\% fewer tokens than $L_3$ yet achieves the highest pooled $F_1$ for all three models: 84.49\% for Qwen, 77.12\% for OLMo, and 89.20\% for DeepSeek.
For a reverse engineer, the choice of setting depends on the model: for DeepSeek, $L_3$ provides slightly higher recovery scores than $R$.
For similar reasons, $L_1$ seems preferable for Qwen and OLMo.
Fixed retrieval (\ie, setting $R$) consumes more tokens than $A$ without matching its detection scores (Figure~\ref{fig:detection-resource-quality}a).
Qwen and DeepSeek consume about 3.4 and 5.0 times as many tokens under $R$ as under $A$, yet achieve lower detection $F_1$: 34.94\% and 66.13\%, respectively.
For OLMo, $R$ reaches 65.41\%, almost the same as the cheaper $L_1$ setting at 65.52\%.
Agentic inspection therefore seems preferable for detection when both accuracy and token cost matter.
The retrieval setting also comes at the cost of preprocessing: construction of the dataset used 315.6 million UniXcoder embedding tokens in just under one hour on our H100 GPU.

\textit{Inference time.}
These runs show no consistent recovery benefit from greater latency; the relation between time and quality depends strongly on which model we use.
Figure~\ref{fig:semantic-recovery-resource-quality}-(c--d) shows per-focal-method inference time: this denotes client-observed elapsed time, including request handling, server queueing, and generation; DeepSeek additionally includes remote API communication, while $A$ also includes retrieval and workflow overhead.
The figure shows that Qwen's mean inference time rises from 1.34\,s per focal-method under $L_0$ to 1.99\,s under $L_3$, alongside a BERTScore $F_1$ increase from 53.07\% to 55.64\%.
OLMo's corresponding time increases more sharply, from 2.39\,s to 6.53\,s, for a smaller BERTScore $F_1$ gain of 1.71 percentage points.
The highest recovery scores come from DeepSeek. In the recorded API runs, it needs 9.47\,s per focal-method for $L_0$, rising to 30.62\,s for $L_3$.
Agentic recovery takes 2.50\,s for Qwen, 4.80\,s for OLMo, and 48.42\,s for DeepSeek, yet none exceeds its own $L_3$ BERTScore $F_1$ or ROUGE-L $F_1$.
The latency of agentic detection depends strongly on the model: $A$ takes 3.74\,s per focal-method for Qwen, 7.29\,s for OLMo, and 69.85\,s for DeepSeek (Figure~\ref{fig:detection-resource-quality}b).
For DeepSeek, $R$ returns results in 6.01\,s, but its detection $F_1$ is 66.13\%, compared with 89.20\% under $A$.

\textit{Traditional-baseline costs.}
Baseline costs vary greatly and must be reported according to each tool's input unit since methods, Java files, JARs, and APKs are not directly comparable.
The following preprocessing times were recorded on our Intel Xeon Phi 7230 machine.
For each fold, Isobe--Tamada needed about 4.8\,s to train its random forests.
Anti-ProGuard required 4\,h 51\,min to extract the fingerprint database from \num{1684} APKs, plus about 15\,min for SQLite indexing. We then spent eight hours evaluating 150 APKs, using 48 workers in total.

For our DeGuard reimplementation, a training pass through \num{1684} application graphs cost 33.85 hours, excluding extraction of the graphs. Each application took another 2\,h 48\,min to process.
No extra training was needed for the released DOBF checkpoint. Its CPU runs nevertheless accumulated 39.70--47.65 hours per obfuscator, split across twelve shards. The longest shard took 3.52--4.10 hours.
These timings make application-wide Anti-ProGuard and DeGuard, and DOBF, costly in our measurements.

\bluebox{\textbf{\hypertarget{aq:3}{\hyperlink{rq:3}{AQ3}}:} Richer static context improves semantic recovery overall, with larger gains for DeepSeek than for Qwen and OLMo, while increasing token consumption for every model.
Considering only raw performance can be detrimental: $L_2$ approaches DeepSeek's $L_3$ recovery quality while using approximately 14\% fewer tokens.
Agentic inspection $A$ gives the strongest detection results and uses fewer tokens than $L_3$, but yields lower semantic-recovery scores for every model.
Qwen has the lowest mean inference time, whereas traditional baselines are generally extremely slow.
The context most useful for identifying obfuscated names is therefore not necessarily the most useful for reconstructing them; the choice depends on the task, model, and inference budget.}

\section{Ablation Study}
\label{sec:ablation}

We digress on some alternative experimental choices that we considered but did not pursue in our main evaluation.
We use two cohorts in this section: a 5\% subset of each obfuscator to cover the full range of target counts while preserving the overall distribution (this totals 8,771 focal-methods); and a 100-focal-method Allatori subset, selected to approximate 100 linearly spaced target-count points between 1 and 64. 
The use of such auxiliary datasets is summarized in \autoref{tab:experiment-campaign-status}.

{
\begin{table*}[t]
  \centering
  \caption{{Cohorts used for experiment campaigns of our ablation studies. For the main experiment campaigns, see \autoref{tab:identifier_target_stats}. $I$=identifier-renaming; $S$=string encryption; $C$=control-flow obfuscation.}}
  \label{tab:experiment-campaign-status}
  \footnotesize
  \setlength{\tabcolsep}{5pt}
  \renewcommand{\arraystretch}{1.14}
  \begin{tabularx}{\textwidth}{@{}>{\raggedright\arraybackslash}p{0.11\textwidth}>{\raggedright\arraybackslash}p{0.20\textwidth}>{\raggedright\arraybackslash}p{0.31\textwidth}>{\raggedright\arraybackslash}X@{}}
    \toprule
    \textbf{Task} & \textbf{Model} & \textbf{Conditions} & \textbf{Cohort} \\
    \midrule
    Recovery & Qwen + Codex & Optional/required retrieval & Stratified 5\% \\
    Recovery & Qwen, OLMo & End-to-end $A$ & Stratified 5\% \\
    Both & Qwen, OLMo & Allatori $I$, $IS$, $IC$, $ISC$ & 100 paired methods each \\
    \bottomrule
  \end{tabularx}
\end{table*}
}

{
\subsection{Harness comparison}
On the stratified 5\% cohort, we compared our $A$ harness with Codex CLI~\cite{openai2025codex}.
Codex CLI is OpenAI's open-source coding agent, which runs in the terminal and can inspect code, modify files, and execute commands.
It forms part of the broader Codex platform that has gained substantial popularity among developers and non-technical users.
Both harnesses allow at most two read-only occurrence inspections and 700 evidence tokens; Codex was tested with optional or required retrieval.
In setting $A$, our task-specific harness asks Qwen which identifier occurrences to inspect, then retrieves the requested evidence, and then asks Qwen to recover the names. 
In the Codex optional-retrieval condition, Codex may request up to two inspections but can proceed without retrieval. 
The required-retrieval version cannot make a final prediction until Codex has requested at least one inspection.

Our agent uses 21.90M tokens and reaches 3.13\% ExM and 9.93\% RL-$F_1$. Required-retrieval Codex uses 115.67M tokens for scores of 2.57\% and 7.73\%, respectively (Table~\ref{tab:codex-retrieval-ablation}).
Our compact task-specific message sequence takes fewer tokens than the general-purpose agent instructions Codex sends with each planning and recovery call. Both also include the task prompt. 
Among the Codex wrappers are \texttt{<system\_message>}\ldots\texttt{</system\_message>}. Its directives include ``Execute the following task'' and ``Return only the JSON in the required schema.''
These wrappers and directives are sent to the LLM twice, once in each pass. They account for most of the input-token overhead. Since the two systems produce outputs of similar length, the excess consumption comes almost entirely from input. Our setup therefore favors the task-specific controller over the more elaborate harness.

\suppressfloats[t]
{
\begin{table}[t]
  \centering
  \caption{{Harness comparison with Qwen on the same stratified 5\% cohort (8,771 methods).
  Scores are method-averaged percentages; tokens and cumulative workflow time include all three obfuscators.
  Time is summed per-method latency, not campaign wall time.}}
  \label{tab:codex-retrieval-ablation}
  \scriptsize
  \setlength{\tabcolsep}{3.3pt}
  \renewcommand{\arraystretch}{1.08}
  \begin{tabular*}{\linewidth}{@{\extracolsep{\fill}}llrrrrr@{}}
    \toprule
    \textbf{Harness} & \textbf{Policy} & \textbf{ExM} & \textbf{RL-$F_1$}
    & \textbf{B$F_1$} & \textbf{Tokens} & \textbf{Time (h)} \\
    \midrule
    Task-specific & $A$ & \textbf{3.13} & \textbf{9.93} & \textbf{48.96} & 21.90M & 18.80 \\
    \midrule
    Codex & Optional retrieval & 2.55 & 7.65 & 46.88 & 115.74M & 27.45 \\
          & Required retrieval & 2.57 & 7.73 & 47.09 & 115.67M & 25.79 \\
    \bottomrule
  \end{tabular*}
\end{table}
}

\subsection{End-to-end detection and recovery}
\label{ssec:end-to-end-recovery}
{On the 5\% cohort of focal-methods from Allatori, yGuard and ProGuard, we run the agentic semantic recovery from each model's saved $A$ detections $\widehat{U}(m)$ from \autoref{ssec:detection-results}, including possible false positives, instead of supplying the ground-truth occurrence inventory $U(m)$.}
Detection $F_1$ reaches 63.17\% for Qwen and 57.04\% for OLMo.
{The resulting method-macro end-to-end BERTScore $F_1$ is 44.14\% for Qwen and 42.92\% for OLMo, compared with 52.58\% and 47.62\%, respectively, for oracle-target $A$ recovery on the same identities with the same aggregation.}
{For this comparison, oracle-target $A$ recovery is restricted to the same target identities as the end-to-end evaluation; Table~\ref{tab:semantic-recovery-pooled} reports results for the full cohort.}
{Within each focal-method, occurrence-level similarity scores are first averaged for each recovered true target. All true targets remain in the semantic-scoring denominator. Targets missed by the detector, or selected targets without a valid recovery, receive zero. We then average over all true targets in each method and over focal-methods with equal weight, following the method-macro aggregation of \autoref{ssec:semantic-recovery-results}. Semantic-recovery scores evaluate genuine targets only; false-positive selections are penalized by detection $F_1$.}

{Names recovered end to end from $\widehat{U}(m)$ score below names recovered from ground-truth occurrences $U(m)$, especially on exact and lexical similarity.}
Both models lose roughly half their exact matches. Their ROUGE-L $F_1$ scores fall by 48--54\%.
The decrease for BERTScore $F_1$ is smaller (10--16\%). With targets selected by the model, the recovered names thus seem less faithful to the originals while often retaining their general meaning.

\subsection{Compounding obfuscations}

In this ablation, we use Qwen and OLMo to study the effects of compounding obfuscations with the other capabilities of Allatori~\cite{allatoriDocumentation}.
We use string encryption (\(S\)) to replace readable constants with encrypted strings and runtime decryption logic. In Allatori, it is configured using \texttt{string-encryption}, \texttt{string-encryption-type}, and \texttt{string-encryption-version} parameters, which we all keep to their default values, \texttt{fast} and \texttt{v4}: this corresponds to Allatori's proprietary encryption algorithm. The call now uses \texttt{log(decrypt(ciphertext))} in place of \texttt{log("OK")}. The added logic decrypts the message at runtime.

Control-flow obfuscation (\(C\)) makes the execution paths in a method body harder to follow by rewriting them, while preserving the program behavior.
It is configured using \texttt{control-flow-obfuscation} and \texttt{extensive-flow-obfuscation}.
We set the extensive-flow level to \texttt{normal}, its default. One possible $C$ rewrite of a simple Java \texttt{if-else} uses nested conditionals and indirect jumps that return the same value.
$I$ denotes renaming: this is the obfuscation we've been considering throughout this work; Allatori replaces meaningful class, method, and field names with short, uninformative names.
We denote renaming with hidden strings by $IS$, renaming with rewritten control flow by $IC$, and the combination of all transformations by $ISC$.

Once the compiled projects are obfuscated using these parameters, we associate the obfuscated focal-methods, obtained through $I$, to their $IS$, $IC$, and $ISC$ variants.
Evaluating on $I$ is exactly the $L_0$ setting we've considered in this paper.
Similarly, we assume Qwen and OLMo only have access to the obfuscated methods: this is still an $L_0$ setting, except the input focal-methods underwent one or several obfuscations.
{As in the task split in \autoref{ssec:task-definition}, Qwen and OLMo are evaluated separately on detecting target occurrences $U(m)$ and recovering names from the true occurrence inventory $U(m)$, with scoring over distinct targets in $T(m)$.}
Each condition in Figure~\ref{fig:allatori-factorial-comparison} contains the same 100 unique methods in this setup.

\begin{figure*}[t]
  \color{black}
  \centering
  \includegraphics[width=\textwidth]{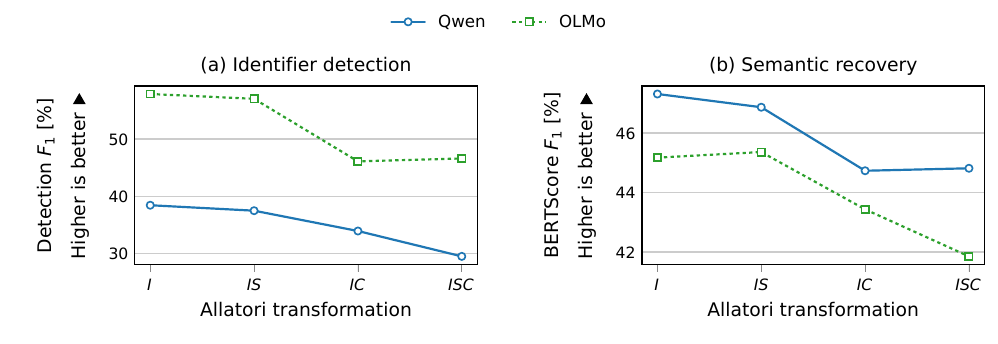}
  \caption{Paired Allatori factorial comparison over 100 focal-methods per condition. Lines link the results of one model for $I$ (renaming alone), $IS$ (added string encryption), $IC$ (added control-flow obfuscation), and $ISC$ (both transformations). Detection $F_1$ is type-agnostic; semantic-recovery B$F_1$ uses oracle targets and method-macro aggregation.}
  \label{fig:allatori-factorial-comparison}
\end{figure*}

Figure~\ref{fig:allatori-factorial-comparison}-(a) shows string encryption cumulated with renaming decreases detection $F_1$ by less than one point with respect to renaming alone $I$ for both models: Qwen's score falls from 38.40\% to 37.45\%, and OLMo falls from 57.86\% to 57.02\%.
The largest loss in detection $F_1$ would be caused by control-flow obfuscation.
Qwen's score falls to 33.89\% on $IC$ and 29.46\% on $ISC$; OLMo reaches 46.08\% and 46.58\%, respectively.
Semantic recovery results are shown in Figure~\ref{fig:allatori-factorial-comparison}-(b): it exhibits little change under $IS$: Qwen B$F_1$ goes from 47.31\% to 46.87\%, while OLMo rises from 45.17\% to 45.36\%.
When obfuscated with $IC$, these scores decrease to 44.73\% and 43.43\%, respectively; under $ISC$, Qwen reaches 44.81\% and OLMo 41.85\%.

The effects of combining the transformations are not perfectly additive. 
Going from $IC$ to $ISC$ costs Qwen about 4.4 percentage points of detection $F_1$, while OLMo gains about 0.5 percentage points. Qwen B$F_1$ barely changes; OLMo loses a further 1.58 points. Control-flow rewriting causes the main degradation in these results. Its interaction with string encryption varies with the model and metric.

\bluebox{\textbf{\hypertarget{aq:4}{\hyperlink{rq:4}{AQ4}}:} In our setup, the task-specific harness recovers names better and uses fewer tokens than Codex.
The end-to-end pipeline produces less faithful names than recovery with the correct targets supplied, as would be expected.
String encryption has little effect, whereas control-flow obfuscation makes both detection and naming harder.
Combining these transformations does not always cause a further loss, but remains the harsher obfuscation for the local LLMs.
}
}

\section{Discussion}
\label{sec:discussion}

\textit{Recovering the ground-truth.}
{The advantage of LLMs in \hyperlink{aq:1}{AQ1} supports using them for suggesting names during reverse engineering, but does not confirm they are a reliable reconstruction of the original source; for example, the highest exact-match score is still only 16.42\% but the BERTScore $F_1$ reaches over 60\% (Table~\ref{tab:semantic-recovery-pooled}).
Nonetheless, ExM is a fragile metric in comparison to BERTScore $F_1$ and should not be regarded as a strong indicator for deobfuscation relevance: we find many instances of identifiers where the predicted lexeme was almost identical to the ground-truth (\eg, in \texttt{double memoryUsagePercent = usage.B()}, Qwen would find \texttt{getTotalAllocatedMemory} instead of \texttt{getTotalMemoryAllocated}) or outright exhibited better syntax than the ground-truth: for example, inside \texttt{boolean bl = b.b.b.e.d.c.w(n2, n3, n10, n8, n9, 20); int n11 = bl ? 0x40FFFFFF : 0x25FFFFFF;}, the original name of \texttt{w} is \texttt{isOver} and the prediction is \texttt{isMouseOver}. 
The predicted lexeme would be preferred here: it indicates that the predicate checks whether the mouse pointer is inside a button's bounds, and its result controls the button's hover color.

\textit{Using the two tasks disjointly.}
Conversely, detection reaches as high as 91.56\% $F_1$ (Table~\ref{tab:detection_scores_method_macro}).
Evaluating both tasks separately allows us to identify the performance bottleneck to provide the reverse engineer with the better tool for the intended task.
Given the right setting drawn close to those exposed inside \autoref{ssec:settings}, the software specialist can therefore choose to use the LLMs to confidently spot the suspect identifiers, and then use the semantic recovery to suggest a name for the identifier as a helper tool.
The latter task can be used disjointly from the former, though: the user would simply mark the identifiers that are suspicious, and then feed them to the LLMs to get a list of suggestions.
{Besides, near-perfect target recall is likely not required to preserve semantic recovery scores: we find that, when each model is provided with a subset of the occurrence inventory drawn from $U(m)$ under setting $A$, with a coverage of 25\%, 50\%, 75\%, or 100\% measured over distinct targets in $T(m)$ (we evaluate on the 5\% cohort of 8,771 focal-methods across all three obfuscators), BERTScore $F_1$ on the supplied targets varies by only 0.60 percentage points for Qwen and 0.15 for OLMo (see \autoref{fig:target-coverage-recovery}).}
These results concern random subsets of genuinely obfuscated targets, without false positives (in contrast, our end-to-end evaluation in \autoref{ssec:end-to-end-recovery} includes both false positives and false negatives as it supplies the detector's actual selections).
The results mean, in particular, that when the model is given the focal-method with agentic retrieval, it doesn't benefit much from knowing the complete set of obfuscated identifiers when suggesting names for those already supplied.
Such a statement regarding general manual accuracy is not possible from our conducted experiments but we envision a human study to confirm this hypothesis.}

\textit{Using the LLMs in other ways.}
Users don't need to restrict themselves to recovering only identifiers selected by the detector; manually identified obfuscated targets may also yield useful naming suggestions.
During our complete campaigns, identifiers missed by the detector receive similar average recovery scores to those it finds when both are explicitly supplied as recovery targets (see \autoref{fig:detection-semantic-recovery-link}).
A reverse engineer could consequently still obtain useful suggestions for a suspicious identifier that was overlooked by the detector, by manually adding it to the recovery target lists without the quality being overly impacted by the detection performance.
LLMs can also complement existing deobfuscators: Appendix~\ref{app:composing-deobfuscators} reports a small experiment highlighting benefits of pairing LLMs with Isobe--Tamada.

\textit{Choosing context under a budget.}
A setting as described in \autoref{ssec:settings} is either imposed or chosen, and it is ultimately the choice of the user to strike a balance between time, quality and token cost.
For repeated queries on a large repository, building a retrieval index ($R$) may therefore be worthwhile: for DeepSeek, $L_3$ uses slightly more tokens per method than $R$ for a gain of 0.66 percentage points in BERTScore $F_1$ (see \autoref{fig:semantic-recovery-resource-quality}).

\textit{Training-data exposure.}
Recent repository creation/commit dates and provenance checks reduce contamination risk to an extent, but a newly created repository can still reuse older code.
Our exact-file and syntax-normalized method audit (see \autoref{ssec:data-collection}) removes detected overlap with the disclosed OLMo training corpus; it does not rule out all prior exposure.
The same audit is unavailable for Qwen and DeepSeek because their training corpora are not disclosed, and the selected repositories do not necessarily postdate DeepSeek's release.
Reported scores may therefore reflect some differences in prior exposure.

\section{Related Work}
\label{sec:related-work}

\textit{Identifier renaming and deobfuscation.}
While our study concerns deobfuscation, renaming assistance can also be applied to code that was never obfuscated in the first place.
\citet{ZhangEtAl2023IdentifierRenaming} identify names that need revision using a random-forest classifier and suggest replacements from related code entities and renaming histories.
Their approach addresses naming inconsistencies during software maintenance and needs the entire clear codebase to function, which is a strong hypothesis for obfuscated code.
For deliberately obfuscated applications, we refer the reader to \autoref{ssec:traditional-baselines} where we cite several traditional deobfuscators we compare against.

\textit{Obfuscation-based pretraining.}
\citet{DBLP:conf/iclr/PaulYGKG25} introduce ObscuraCoder, which uses both the obfuscation and deobfuscation of source code as pretraining tasks to improve general-purpose code models (including for Java).
Unlike our two-task decomposition of the identifier recovery problem, ObscuraCoder reconstructs code in a single generation step, without separately detecting obfuscated identifiers; in fact, its training inputs use typed placeholders that explicitly signal identifier replacement, as in VarBERT~\cite{DBLP:conf/sp/PalBBDNBGSASDBW24}, whereas our detection task requires distinguishing obfuscated identifiers from legitimate ones in decompiled Java code.
Their evaluation measures downstream coding capabilities rather than identifier-recovery accuracy and does not report any of the relevant metrics pertaining to deobfuscation; besides, their related-work discussion omits several established identifier-deobfuscation approaches such as VarBERT or DeGuard~\cite{DBLP:conf/ccs/BichselRTV16}.
Our study instead evaluates how various context and retrieval settings affect recovery performance.
ObscuraCoder's comparison with DOBF also uses newly trained decoder-only adaptations, whereas we evaluate the original released encoder-decoder checkpoint.
Although the authors release training code and their obfuscator, we could not recover public pretrained checkpoints for either ObscuraCoder or their adapted DOBF models.

\textit{Name recovery from stripped binaries.}
VarBERT~\cite{DBLP:conf/sp/PalBBDNBGSASDBW24} adapts a pretrained BERT model to predict variable names in decompiled functions and reports improvements in human comprehension.
GENNM~\cite{DBLP:conf/ndss/Xu0SH0Y0X00025} uses generative models to recover variable names and iteratively incorporates predictions from callers and callees.
Similarly, \textsc{SYMGEN} uses LLMs to recover function names in stripped binaries~\cite{DBLP:conf/ndss/JiangJ025}, whereas we evaluate both detection and recovery for class, method, and field identifiers in obfuscated Java with varying context allowances.
These approaches study native binaries with stripped symbols.
In our work, we instead examine deliberate identifier obfuscation in Java, where meaningful and obfuscated names coexist, and additionally compare field, method, and class identifiers across several context/retrieval settings; GENNM also always assumes the availability of the entire program, which is the strongest assumption we make in our study.
Stripping symbols is not equivalent to obfuscation: other than the chosen replacement vocabulary that can change recovery difficulty, the available identifier information depends on the compilation model and binary format. The cited study focuses on variable names.

\textit{Code cleaning and repository context.}
\citet{JainEtAl2024CodeCleaning} use LLMs to clean training code, including renaming variables that are not obfuscated.
Their renaming model receives the problem statement explaining what the program should do.
CatCoder~\cite{PanEtAl2025CatCoder} uses a function's name paired with available documentation and related repository code to write its missing body.
In our tasks, the body already exists, but some names no longer explain what the code does.
We therefore ask models to identify obfuscated names and suggest meaningful replacements while using additional code that is also obfuscated.
CatCoder evaluates whether the generated code compiles and passes tests (conversely, we do not guarantee that the renamed code compiles) whereas our focus is the quality of the names themselves and whether they are detected or not.

\textit{Assistance to reverse engineers.}
\citet{BasqueEtAl2026Synergy} study human--LLM collaboration in reverse engineering.
Their experiment with 48 participants uncovers remarkable comprehension gains for novices but limited improvements for experts, alongside problems caused by hallucinations and unhelpful suggestions.
Our study is a complement to such a work: it examines when models detect obfuscated identifiers and recover meaningful names, but establishing how these suggestions affect reverse-engineering efforts still requires human evaluation.

\section{Conclusion}
\label{sec:conclusion}

By studying identifier renaming in semantically rich obfuscated source code through the lens of LLMs, we address an important gap in the software deobfuscation literature. 
Identifier deobfuscation of focal-methods is best studied as two problems that reward different context and retrieval settings: obfuscated identifier detection first, followed by semantic name recovery. 
Through an evaluation of three LLMs, Qwen, OLMo (which allows for contamination control), and DeepSeek, on 175,864 focal-methods from 3 obfuscators (Allatori, ProGuard, and yGuard), we find that 
read-only agentic inspection is consistently strongest for flagging which visible identifiers are obfuscated, while deterministic enclosing-class and one-hop call context is strongest for recovering a semantically appropriate name. 
Our ablation study favors the use of a simple task-specific harness, and tempers expectations for end-to-end identifier deobfuscation.
The results overall motivate that strength in one task does not imply strength in the other, and notably that raw performance should also be appreciated in light of the time and token cost of each setting, guiding a reverse engineer to choose the best tool under the right circumstance.

Future work should evaluate resilience against dictionary-based renaming and long, high-entropy identifiers, and address the difficulty of class-identifier detection and recovery and the degradation caused by control-flow obfuscation.
A human evaluation is also needed to establish whether partial semantic recovery speeds up reverse-engineering tasks.

\section*{Acknowledgments}
We thank Jacopo Caratti for proposing the original idea behind this paper and for his contributions to earlier versions.

\begin{appendices}
\renewcommand{\theHfigure}{appendix.\arabic{section}.\arabic{figure}}
\renewcommand{\theHtable}{appendix.\arabic{section}.\arabic{table}}

\section{\texorpdfstring{$L_3$}{L3} caller/callee evidence}
\label{app:l3-caller-callee-evidence}

Resolution for $L_3$ uses the owner and signature as well as the name. An unresolved invocation, or one whose target is outside the obfuscated artifact, does not create an edge in the call graph.

Using all available verified one-hop caller and callee bodies, we obtain the focal-method's context as \(L_3 = L_2 + \text{verified caller/callee method bodies, when available}.\)

\subsection{\texorpdfstring{$L_3$}{L3} context construction}
 For curated method $m$, $L_3$ combines the $L_2$ owner-class context with verified caller/callee bodies from the same obfuscation run.
 We build context from the CFR-decompiled Allatori-, ProGuard-, or yGuard-obfuscated Java artifacts alone, without clean source or private mappings. Each repository $R$ has a method call graph
 \(G_R=(V_R,E_R),\)
 whose node $v \in V_R$ identifies an obfuscated method by repository, owner class, name, and bytecode/decompiled signature.
 
We add a directed edge
 \((u,v) \in E_R\)
 for an invocation in the body of method $u$ that resolves to method $v$ in the same obfuscated repository artifact.

These calls define the caller and callee sets of focal-method $m$:
 \[
 \mathrm{Callers}(m)=\{u : (u,m)\in E_R\},
 \qquad
 \mathrm{Callees}(m)=\{v : (m,v)\in E_R\}.
 \]
 
We serialize $L_3$ in the following form:
 \[
 \begin{aligned}
 L_3(m)=&
 \{\mathrm{Focal}(m), \mathrm{Class}(m)\}\\
 &{}\cup
 \{\mathrm{Caller}(u) : u\in \mathrm{Callers}(m)\}\\
 &{}\cup
 \{\mathrm{Callee}(v) : v\in \mathrm{Callees}(m)\},
 \end{aligned}
 \]
 where $\mathrm{Class}(m)$ is the full obfuscated owner-class context already supplied by $L_2$.
 The prompt has a bare \texttt{<focal\_method>} block, then an \texttt{<enriching\_context>} block. The latter contains the owner \texttt{<class\_context>} first, followed by one \texttt{<caller\_context>} block per verified caller and one \texttt{<callee\_context>} block per verified callee.

If no verified caller or callee is available, it is simply not included and $L_3$ degenerates to $L_2$ (\ie, owner-class context as described in \autoref{fig:static-context-levels}).

Among the 175,864 evaluated focal-methods, 109,008 (62.0\%) contain caller/callee evidence; these nonempty samples contain a median of two and a mean of 3.44 method bodies, with at most 446.
We counted the bodies in the saved $L_3$ contexts before truncating them for the prompt.

\begingroup
\renewcommand{\theHtable}{appendix.l3.\arabic{table}}
\begin{table}[t]
    \caption{Availability and composition of caller/callee evidence in $L_3$.
    {Counts cover the evaluated cohort of 175,864 methods pooled across Allatori, ProGuard, and yGuard.}}
    \label{tab:l3_caller_callee_availability}
    \centering
    \small
    \begin{tabular}{@{}lr@{}}
        \toprule
        \textbf{Case} & \textbf{Samples} \\
        \midrule
        Total samples                         & {175,864} \\
        No verified caller/callee             & {66,856} \\
        At least one caller/callee             & {109,008} \\
        \quad Only callers                   & {45,650} \\
        \quad Only callees                   & {26,748} \\
        \quad Both callers and callees     & {36,610} \\
        Multiple caller/callee bodies          & {66,808} \\
        \botrule
    \end{tabular}
\end{table}
\endgroup

\section{Partial replication of obfuscator prevalence}
\label{app:obfuscator-replication}

For this replication, we used the authors' feature-extraction code~\cite{NSSUSYDObfuscationLargeAnalysis} and retrained the classifiers instead of using their released pretrained models. Their repository does not provide the original APK dataset, so we used a snapshot of the AndroZoo metadata~\cite{allix2016androzoo}. This is a qualitative partial replication of predicted prevalence. We have not verified the APKs' build provenance, so the predictions do not establish which obfuscators were used.

The clean APKs came from F-Droid. We identified \(4{,}330\) valid repositories there and used a crawler to explore, verify and compile the projects. For reliable compilation, we limited the builds to native Android applications using Gradle, with either Groovy or Kotlin build files. Before compilation, the pipeline turned off minification and obfuscation by writing \texttt{minifyEnabled false} in Groovy files or \texttt{isMinifyEnabled = false} in Kotlin files. The pipeline compiled the projects and signed the resulting APKs, producing the Unobfuscated Dataset of \(451\) clean APKs.

We extracted all 37 DEX-level features from the authors' list, including the distributions of class, method, field and string lengths. The complete list is in the original study. There were four models to retrain: a multilayer perceptron for obfuscation detection and three random forests for tool identification. Each forest was trained one-vs-rest, with ProGuard, Allatori or DashO as its positive class. ProGuard is free and open-source, while we have academic licenses for Allatori and DashO. We stratified the data into training and validation partitions, using an \(80/20\) ratio and a fixed random seed. For comparison, the authors' \(80/20\) split was accompanied by three-fold cross-validation for hyperparameter tuning. The binary model's accuracy after retraining was \(93.4\%\), with \(92.8\%\) for \(F_1\); the three forests had results similar to those reported in the original study.

We drew \(1{,}000\) package names randomly from the Google Play part of our AndroZoo metadata snapshot and retained one randomly chosen APK per name. For every APK we could parse, we extracted the 37 features and ran the binary classifier. The three tool classifiers were then run on the APKs predicted to be obfuscated. For a tool label, we kept the original rule. We chose the highest probability among the three classifiers; if all were below \(0.5\), the APK was labeled \emph{Other}. The comparison with Niroshan et al.~\cite{niroshan2025empirical} is in Table~\ref{tab:obfuscation_prevalence_comparison}.

\begingroup
\renewcommand{\theHtable}{appendix.obfuscation.\arabic{table}}
\begin{table}[t]
    \caption{Comparison of predicted obfuscation prevalence with the large-scale study of Niroshan et al.~\cite{niroshan2025empirical}.}
    \label{tab:obfuscation_prevalence_comparison}
    \centering
    \small
    \begin{tabular}{@{}lcc@{}}
        \toprule
        \textbf{Metric} & \textbf{Our sample} & \textbf{Niroshan et al.} \\
        \midrule
        Valid APKs                         & \(1{,}000\)           & \(548{,}967\) \\
        Classified as obfuscated           & \(563\) (\(56.3\%\))  & \(308{,}782\) (\(56.25\%\)) \\
        \midrule
        \multicolumn{3}{@{}l}{\textit{Predicted tool among obfuscated APKs}} \\
        \quad ProGuard                     & \(220\) (\(39.1\%\))  & \(40.92\%\) \\
        \quad Allatori                     & \(198\) (\(35.2\%\))  & \(36.64\%\) \\
        \quad DashO                        & \(34\) (\(6.0\%\))    & \(1.01\%\) \\
        \quad Other                        & \(111\) (\(19.7\%\))  & \(21.43\%\) \\
        \bottomrule
    \end{tabular}
\end{table}
\endgroup

ProGuard and Allatori were the two most frequently predicted obfuscator classes, as in the original study. The overall distributions were similar despite the small differences in training and dataset partitioning.

\section{{Type-aware semantic-recovery scores}}
\label{app:semantic-recovery-by-kind}

Table~\ref{tab:semantic-recovery-by-kind} separates exact-match and lexical-similarity scores by identifier type.

\begingroup
\renewcommand{\theHtable}{appendix.semantic.\arabic{table}}

{\color{black}
\begin{sidewaystable}[!p]
    \color{black}
    \centering
    \caption{{Type-aware semantic recovery scores pooled across obfuscators.
    {The upper block gives method-macro LLM scores for $L_0$--$L_3$, $R$, and $A$. Each evaluated focal-method with targets of the reported type has equal weight.} The lower block reports method-macro baseline scores, excluding focal-methods without any emitted prediction; unsupported or unobserved types are marked \quotes{--} (Anti-ProGuard doesn't rename fields and Isobe-Tamada doesn't touch F and C). F, M, and C denote fields, methods, and classes; scores are percentages.}}
    \label{tab:semantic-recovery-by-kind}
    \scriptsize
    \setlength{\tabcolsep}{2.5pt}
    \renewcommand{\arraystretch}{1.02}
    \begin{tabular*}{\textheight}{@{\extracolsep{\fill}}ll*{12}{r}@{}}
        \toprule
        \textbf{Setting} & \textbf{Model}
        & \multicolumn{3}{c}{\textbf{ExM}}
        & \multicolumn{3}{c}{\textbf{LCS}}
        & \multicolumn{3}{c}{\textbf{Jac}}
        & \multicolumn{3}{c}{\textbf{RL-$F_1$}} \\
        \cmidrule(lr){3-5}\cmidrule(lr){6-8}\cmidrule(lr){9-11}\cmidrule(lr){12-14}
        & & \textbf{F} & \textbf{M} & \textbf{C}
        & \textbf{F} & \textbf{M} & \textbf{C}
        & \textbf{F} & \textbf{M} & \textbf{C}
        & \textbf{F} & \textbf{M} & \textbf{C} \\
        \midrule
        $L_0$ & Qwen & 12.71 & 2.89 & 2.40 & 35.43 & 32.97 & 31.59 & 16.88 & 13.51 & 8.83 & 18.38 & 18.76 & 11.65 \\
         & OLMo & 9.26 & 2.75 & 2.79 & 31.32 & 30.72 & 29.74 & 13.83 & 12.77 & 9.49 & 15.44 & 17.53 & 12.38 \\
         & DeepSeek & 17.74 & 7.00 & 4.97 & 39.14 & 36.33 & 32.29 & 22.84 & 20.35 & 13.87 & 24.49 & 26.50 & 17.54 \\
        \midrule
        $L_1$ & Qwen & 16.38 & 4.09 & 2.91 & 39.06 & 34.13 & 33.29 & 21.22 & 15.01 & 11.15 & 22.90 & 20.28 & 14.83 \\
         & OLMo & 13.39 & 3.52 & 3.11 & 33.50 & 30.81 & 30.77 & 18.03 & 13.13 & 10.76 & 19.62 & 17.69 & 14.02 \\
         & DeepSeek & 22.35 & 10.61 & 6.48 & 42.92 & 40.28 & 35.42 & 27.97 & 24.85 & 17.58 & 29.81 & 31.24 & 22.11 \\
        \midrule
        $L_2$ & Qwen & 17.45 & 4.96 & 3.19 & 39.88 & 35.37 & 34.39 & 22.51 & 16.44 & 12.23 & 24.28 & 21.89 & 16.23 \\
         & OLMo & 13.93 & 3.60 & 2.70 & 33.81 & 30.95 & 30.92 & 18.41 & 12.92 & 10.05 & 19.96 & 17.36 & 13.27 \\
         & DeepSeek & 24.50 & 13.18 & 7.55 & 45.37 & 43.21 & 38.34 & 30.89 & 28.43 & 20.37 & 32.98 & 35.03 & 25.54 \\
        \midrule
        $L_3$ & Qwen & 17.25 & 5.12 & 3.51 & 39.95 & 35.95 & 35.02 & 22.45 & 17.05 & 12.96 & 24.28 & 22.65 & 17.13 \\
         & OLMo & 13.54 & 3.79 & 2.65 & 33.78 & 31.72 & 31.36 & 18.04 & 13.46 & 10.19 & 19.62 & 18.04 & 13.50 \\
         & DeepSeek & 24.88 & 13.96 & 8.22 & 46.24 & 44.49 & 39.49 & 31.58 & 29.79 & 21.69 & 33.76 & 36.55 & 27.09 \\
        \midrule
        $R$ & Qwen & 12.25 & 3.37 & 2.20 & 34.81 & 33.88 & 32.14 & 16.29 & 14.34 & 9.13 & 17.77 & 19.68 & 12.22 \\
         & OLMo & 9.39 & 2.22 & 2.40 & 30.63 & 30.77 & 31.37 & 13.50 & 11.96 & 8.80 & 14.97 & 16.64 & 11.62 \\
         & DeepSeek & 32.93 & 10.43 & 13.26 & 54.52 & 42.72 & 46.81 & 40.82 & 26.26 & 31.04 & 43.43 & 32.73 & 37.96 \\
        \midrule
        $A$ & Qwen & 13.20 & 3.01 & 2.55 & 35.16 & 32.46 & 31.09 & 17.17 & 12.75 & 9.45 & 18.60 & 17.55 & 12.46 \\
         & OLMo & 10.62 & 3.22 & 2.82 & 29.69 & 29.29 & 26.19 & 14.53 & 13.16 & 8.88 & 15.90 & 17.86 & 11.46 \\
         & DeepSeek & 22.15 & 8.49 & 6.31 & 43.20 & 36.76 & 34.77 & 27.91 & 20.90 & 17.70 & 29.80 & 26.39 & 22.25 \\
        \hline\hline
        \noalign{\vskip 3pt}
        Leading verb & Isobe--Tamada
          & -- & 5.25 & --
          & -- & 22.23 & --
          & -- & 5.25 & --
          & -- & 5.25 & -- \\
        Native & Anti-ProGuard
          & -- & 0.00 & 0.00
          & -- & 27.63 & 16.35
          & -- & 9.03 & 0.45
          & -- & 15.07 & 0.73 \\
        Native & DeGuard
          & 1.78 & 0.67 & 0.01
          & 8.66 & 13.51 & 10.48
          & 2.31 & 3.48 & 0.48
          & 2.42 & 4.80 & 0.68 \\
        Native & DOBF
          & 4.58 & 2.05 & 1.26
          & 17.27 & 16.84 & 14.13
          & 7.19 & 7.06 & 4.12
          & 7.90 & 9.42 & 5.37 \\
        \bottomrule
    \end{tabular*}
\end{sidewaystable}
}

\endgroup
\clearpage

\section{Detection and semantic recovery}
\label{app:detection-semantic-recovery-link}

\begin{figure}[t]
  \centering
  \includegraphics[width=\linewidth]{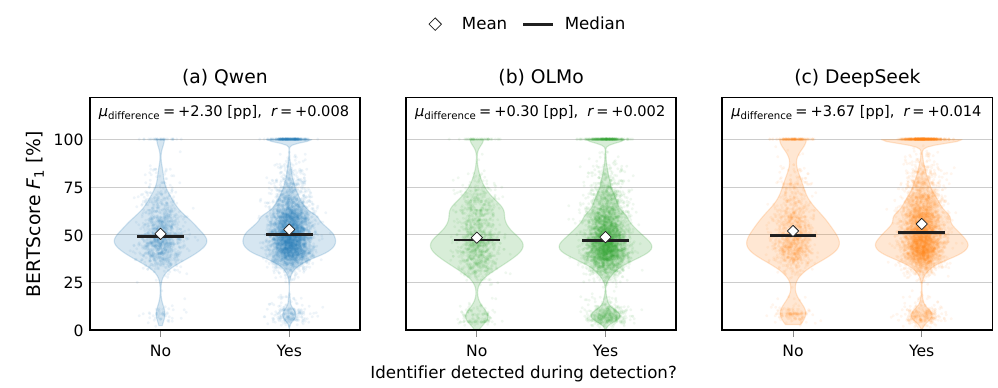}
  \par\vspace{4pt}
  \caption{{Semantic-recovery performance conditioned on whether or not the distinct target was detected (both tasks performed under the $A$ setting).
  {$\mu_{\text{difference}}$ reports the mean difference in B$F_1$ between detected and missed targets, and $r$ is the point-biserial correlation between detection and semantic recovery (0 means no correlation).}}}
  \label{fig:detection-semantic-recovery-link}
\end{figure}

{

{Figure~\ref{fig:detection-semantic-recovery-link} compares semantic recovery for distinct targets selected or missed during agentic detection.}
{Each subplot includes all targets identified by unique private mapping IDs (307,097 targets).}
{Detected targets (TP for detection) obtain slightly higher mean B$F_1$ than missed targets (FN for detection): 52.82 versus 50.52 for Qwen, 55.71 versus 52.04 for DeepSeek, and 48.81 versus 48.51 for OLMo.}
The point-biserial correlations are small ($r=0.002$--$0.014$), so successful detection alone is a weak indicator of naming quality in these outputs.
}

\begin{figure}[htbp]
  \centering
  \includegraphics[width=.5\linewidth]{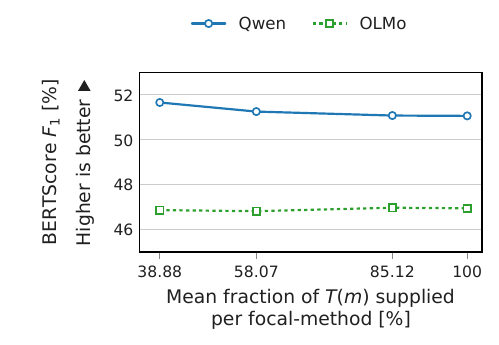}
  \caption{{{Effect of incomplete oracle occurrence inventories drawn from $U(m)$ on semantic recovery (setting $A$) for Qwen and OLMo on 8,771 focal-methods.}
  {Omitted targets are excluded from the scores: each supplied distinct target is scored once.}
  The horizontal axis reports achieved coverage of $T(m)$: 38.88\%, 58.07\%, 85.12\%, and 100\% (these values are obtained after rounding the nominal 25\%, 50\%, 75\%, and 100\% inventories upwards).}}
  \label{fig:target-coverage-recovery}
\end{figure}

\clearpage
\section{Composing deobfuscators}
\label{app:composing-deobfuscators}

We also underline the possibility of using LLMs as complementary tools to existing deobfuscators.
We ran Qwen and OLMo under the agentic recovery setting on three repositories obfuscated with Allatori, ProGuard, and yGuard (which makes nine repository--obfuscator instances per model), and then filled each undetected method-name verb with the corresponding Isobe--Tamada prediction.

On this small cohort, Table~\ref{tab:llm_isobe_pipeline} shows higher recovery scores when an LLM precedes Isobe--Tamada: BERTScore $F_1$ rises from 50.55\% to 57.29\% with Qwen and 54.77\% with OLMo.
ROUGE-L $F_1$ also improves for both models, whereas detection gains remain below one point and SemSim is essentially unchanged.

\newcommand{\updelta}[1]{\textcolor{green!50!black}{(\(\uparrow\)#1)}}
\newcommand{\downdelta}[1]{\textcolor{red!70!black}{(\(\downarrow\)#1)}}

\begin{table}[!htbp]
    \centering
    \caption{We compare Isobe--Tamada alone with Isobe--Tamada preceded by Qwen/OLMo detection and semantic recovery; scores are averages pooled over methods and obfuscators.
    Parentheses show the change in percentage points relative to Isobe--Tamada alone, calculated from the displayed scores.}
    \label{tab:llm_isobe_pipeline}
    \begin{tabular}{lrrr}
        \toprule
        \textbf{Metric}
        & \textbf{Isobe before}
        & \textbf{Qwen $\rightarrow$ Isobe}
        & \textbf{OLMo $\rightarrow$ Isobe} \\
        \midrule
        Detection
        & 98.89\%
        & 99.67\% \updelta{0.78}
        & 99.78\% \updelta{0.89} \\
        B$F_1$
        & 50.55\%
        & 57.29\% \updelta{6.74}
        & 54.77\% \updelta{4.22} \\
        RL-$F_1$
        & 7.62\%
        & 19.91\% \updelta{12.29}
        & 18.33\% \updelta{10.71} \\
        SemSim
        & 99.74\%
        & 99.74\% (0.00)
        & 99.74\% (0.00) \\
        \bottomrule
    \end{tabular}
\end{table}

\end{appendices}

\backmatter

\section*{Data and code availability}
The replication package is publicly available on Figshare~\cite{QueyrutEtAl2026Artifact} at \url{https://doi.org/10.6084/m9.figshare.34572789.v1}.
It includes dataset archives for the static settings and traditional baseline implementations with trained models.

\bibliography{references}

\end{document}